\pdfoutput=1

\documentclass[iop, apj]{emulateapj}

\usepackage{xspace}
\usepackage{amsmath}
\usepackage{framed} 
\usepackage{txfonts}
\usepackage{epstopdf} 
\usepackage{color}
\usepackage{rotating}
\usepackage{natbib}
\usepackage{ulem}
\usepackage{xspace}
\usepackage[colorlinks=true,urlcolor=blue,linkcolor=blue,citecolor=blue]{hyperref}

\slugcomment{To be submitted to the Astrophysical Journal}

\newdimen\hssize
\newcommand{\pc}{\>{\rm pc}}

\newcommand{\mpc}{\>{\rm Mpc}}

\newcommand{\mpch}{\>h^{-1}{\rm {Mpc}}}

\newcommand{\kmsmpc}{\>{\rm km}\,{\rm s}^{-1}\,{\rm Mpc}^{-1}}

\def\gcm3{\mathrm{g} / \mathrm{cm}^3}

\def\gtsima{$\; \buildrel > \over \sim \;$}
\def\ltsima{$\; \buildrel < \over \sim \;$}
\def\prosima{$\; \buildrel \propto \over \sim \;$}
\def\gsim{\lower.7ex\hbox{\gtsima}}
\def\lsim{\lower.7ex\hbox{\ltsima}}
\def\simgt{\lower.7ex\hbox{\gtsima}}
\def\simlt{\lower.7ex\hbox{\ltsima}}
\def\simpr{\lower.7ex\hbox{\prosima}}

\newcommand{\avg}[1]{\langle #1 \rangle}

\def\rmb{{\rm b}}
\def\rmc{{\rm c}}
\def\rmd{{\rm d}}

\def\rmg{{\rm g}}
\def\rmh{{\rm h}}

\def\rmm{{\rm m}}

\def\rmp{{\rm p}}

\def\rms{{\rm s}}

\def\rmx{{\rm x}}
\def\rmy{{\rm y}}

\def\calH{{\cal H}}

\usepackage{etoolbox}
\makeatletter
\patchcmd{\NAT@citex}
  {\@citea\NAT@hyper@{\NAT@nmfmt{\NAT@nm}\NAT@date}}
  {\@citea\NAT@nmfmt{\NAT@nm}\NAT@hyper@{\NAT@date}}
  {}
  {}
\patchcmd{\NAT@citex}
  {\@citea\NAT@hyper@{%
     \NAT@nmfmt{\NAT@nm}%
     \hyper@natlinkbreak{\NAT@aysep\NAT@spacechar}{\@citeb\@extra@b@citeb}%
     \NAT@date}}
  {\@citea\NAT@nmfmt{\NAT@nm}%
   \NAT@aysep\NAT@spacechar%
   \NAT@hyper@{\NAT@date}}
  {}
  {}
\patchcmd{\NAT@citex}
  {\@citea\NAT@hyper@{%
     \NAT@nmfmt{\NAT@nm}%
     \hyper@natlinkbreak{\NAT@spacechar\NAT@@open\if*#1*\else#1\NAT@spacechar\fi}%
       {\@citeb\@extra@b@citeb}%
     \NAT@date}}
  {\@citea\NAT@nmfmt{\NAT@nm}%
   \NAT@spacechar\NAT@@open\if*#1*\else#1\NAT@spacechar\fi%
   \NAT@hyper@{\NAT@date}}
  {}
  {}
\makeatother

\shorttitle{Astrophysical and cosmological constraints from clustering and lensing of BOSS galaxies}
\shortauthors{More et al.}

\begin{document}

%
\def\figdir{.}
\def\figext{pdf}


\title{The weak lensing signal and the clustering of BOSS galaxies II:
Astrophysical and Cosmological constraints}
\author{Surhud More \altaffilmark{1}, Hironao~Miyatake \altaffilmark{2,1},
Rachel~Mandelbaum  \altaffilmark{3}, Masahiro~Takada  \altaffilmark{1},
David.~N.~Spergel  \altaffilmark{2},
Joel~R.~Brownstein  \altaffilmark{4},
Donald P. Schneider \altaffilmark{5,6}
}

\affil{
$^1$ Kavli Institute for the Physics and Mathematics of the Universe
(WPI), TODIAS,
The
University of Tokyo, 
Chiba, 277-8583, Japan \\
$^2$ Department of Astrophysical Sciences, Princeton University, Peyton Hall,
Princeton NJ 08544, USA \\
$^3$ McWilliams Center for Cosmology, Department of Physics, Carnegie Mellon University, Pittsburgh, PA
15213, USA\\
$^4$ Department of Physics and Astronomy, University of Utah, 115 S 1400
E, Salt Lake City, UT 84112, USA\\
$^5$ Department of Astronomy and Astrophysics, The Pennsylvania State
University, University Park, PA 16802, USA\\
$^6$ Institute for Gravitation and the Cosmos, The Pennsylvania State
University, University Park, PA 16802, USA\\
}


\begin{abstract}
We perform a joint analysis of the abundance, the clustering and the
galaxy-galaxy lensing signal of galaxies measured from Data Release 11 of the
Sloan Digital Sky Survey III Baryon Oscillation Spectroscopic Survey (SDSS
III-BOSS) in our companion paper, Miyatake et al. (2014). The lensing signal was
obtained by using the shape catalog of background galaxies from the Canada
France Hawaii Telescope Legacy Survey, which was made publicly available by the
CFHTLenS collaboration, with an area overlap of about 105~deg$^2$.  We analyse
the data in the framework of the halo model in order to fit halo occupation
parameters and cosmological parameters ($\Omega_\rmm$ and $\sigma_8$) to these
observables simultaneously, and thus break the degeneracy between galaxy bias
and cosmology. Adopting a flat $\Lambda$CDM cosmology with priors on $\Omega_b
h^2$, $n_\rms$ and $h$ from the analysis of WMAP 9-year data, we obtain
constraints on the stellar mass-halo mass relation of galaxies in our
sample. Marginalizing over the halo occupation distribution parameters and a
number of other nuisance parameters in our model, we obtain
$\Omega_\rmm=0.310^{+0.019}_{-0.020}$ and $\sigma_8=0.785^{+0.044}_{-0.044}$
(68\% confidence).  We demonstrate the robustness of our results with respect to
sample selection and a variety of systematics such as the halo off-centering
effect and possible incompleteness in our sample.  Our constraints are
consistent, complementary and competitive with those obtained using other
independent probes of these cosmological parameters.  The cosmological analysis
is the first of its kind to be performed at a redshift as high as $0.53$.
\end{abstract}

\keywords{cosmology: theory - cosmology: observations - large-scale structure of
universe - gravitational lensing: weak - cosmology: cosmological parameters}


\section{Introduction}
\label{sec:intro}

The overwhelming majority of the energy density of the Universe today is
dominated by two mysterious components -- dark energy and cold dark matter --
both motivated by astrophysical observations \citep[see e.g,][]{Ostriker:1974, Rubin:1978, Riess:1998, Perlmutter:1999, Hinshaw:2013, Planck:2013}. Since their discovery, the field of
observational cosmology has focused on characterizing the precise abundance, the
statistical distribution and the phenomenological behaviour of these components.
Geometrical probes such as the observations of type-Ia supernovae \citep[see
e.g.,][]{Lampeitl:2010, Sullivan:2011, Suzuki:2012} and the baryonic acoustic
oscillation measurements \citep[see e.g.,][]{Eisenstein:2005, Percival:2007,
Blakeetal:11, Anderson:2014} have provided constraints on the energy density of
various components in the Universe as a function of redshift, but are
insensitive to the statistical properties of the dark matter distribution.
Probing the latter requires constraints on the growth of structure in the
Universe, which can be provided by measurements of the abundance of galaxy
clusters \citep[see e.g.,][]{Vikhlinin:2009, Mantz:2010, Rozo:2010, Benson:2013,
Hasselfield:2013, Planck:2013a}, redshift space distortions \citep[see
e.g.,][]{Percival:2004, Beutler:2012, Reid:2014} and the statistics of weak
gravitational lensing as a function of redshift \citep[see
e.g.,][]{Waerbeke:2000, Lin:2012, Huff:2014, Heymans:2013, Mandelbaum:2013}.
Over the next decade, a combination of these probes will enable a
phenomenological understanding of the nature of dark energy and dark matter as
well as stringent constraints on modifications to gravity \citep[see
e.g.,][]{Albrecht:2006}. 

The growth of structure in the Universe is driven by the growth of fluctuations
in dark matter, which are easier to describe analytically on large scales
\citep{Bernardeu:2002} or via collisionless numerical simulations on small
scales \citep{Davis:1985} than the variety of astrophysical processes that
baryons undergo in order to form galaxies \citep[see
e.g.,][]{Springel:2005,Rudd:2008, Vogelsberger:2014}. Observationally, however, it is
easier to use galaxies to trace out the underlying structure in matter. Since
galaxies form within halos, at the peaks of the matter density field, using
galaxies as tracers produces a biased view of the matter distribution
\citep{Kaiser:1984}. The bias of halos with respect to the matter distribution
and its dependence on halo mass can be fortunately predicted given the
cosmological parameters within the framework of the standard concordance
cosmological model \citep{Bardeen:1986, MoWhite:96, Sheth:1999, Sheth:2001,
Tinker:2010}.

On large scales the bias of halos and the galaxies that reside in them
approaches a constant value. On such scales the shape of the matter two-point
function (the power spectrum or the correlation function) can be inferred from
the observed galaxy two-point function, and used to constrain cosmological
parameters
\citep{Tegmark:2004,Percival:2007a,Reid:2010,Saitoetal:11}. However, in the case
of the galaxy two-point function, the amplitude of the matter power spectrum,
which is essential to study the growth of structure, is entirely degenerate with
the value of the bias. The determination of galaxy bias can be complicated as
it is known to depend upon the properties of galaxies such as their luminosity
and colour \citep{Norberg:2001, Tegmark:2004, Zehavi:2011, Guo:2013}, and is
quite scale dependent on small scales \citep{FryGaztanaga:93,Mann:1998,
Cacciato:2012}.
Nevertheless, this degeneracy between the large scale bias and the amplitude of
the matter power spectrum can be broken if there is a way to infer the
connection between galaxies and their halo masses \citep{Seljak:2005}.

There are a number of different approaches to directly infer the galaxy-dark matter
connection. Many different observables can be used to probe this connection,
including galactic rotation curves \citep{Rubin:1983}, kinematics of
satellite galaxies \citep{Zaritsky:1997,vdBosch:2004,More:2009,More:2011}, 
small scale redshift space distortions \citep{Hikage:2013,Li:2012}, $X$-ray emission
from the hot intra-cluster medium \citep[see reviews
by][]{Kravtsov:2012,Ettori:2013}. However, these methods assume that
the system is in dynamical equilibrium, an assumption that is
certainly violated in some systems.  Weak gravitational lensing provides a way
to circumvent this assumption and can be used as a relatively clean probe of
the halo masses. In combination with weak lensing, the information encapsulated
in the shape and amplitude of the clustering signal can be fully exploited 
\citep{Seljak:2005, Cacciato:2009, Mandelbaum:2013, Hikage:2013,Cacciato:2013, More:2013,
Reid:2014}. This combination can provide simultaneous constraints on the
connection between galaxies and dark matter and the cosmological parameters.
Cosmological constraints from such studies obtained at different cosmic epochs
can then be used to constrain the equation of state of dark energy.

In \citet[][Paper I hereafter]{Miyatake:2013}, we measure the large scale
clustering of galaxies in the Sloan Digital Sky Survey III (SDSS-III hereafter)
Baryon Oscillation Spectroscopic Survey (BOSS hereafter). In particular, we
employ the CMASS galaxy sample from BOSS as our parent sample. We use the deep
but limited area imaging data from the Canada France Hawaii Telescope Legacy
Survey (CFHTLS hereafter), to measure the weak gravitational lensing signal
around galaxies from BOSS to calibrate the masses of the halos in which they
reside. In this paper, we model these observations simultaneously in the
framework of the halo model. We will obtain joint constraints on the
astrophysical properties of galaxies in our sample such as their halo occupation
distribution, limiting constraints on their stellar masses, and the density
profile of dark matter halos in which they reside, as well as cosmological
constraints on the matter density parameter $\Omega_{\rmm}$ and the amplitude of
density fluctuations characterized by the parameter $\sigma_8$.

This paper is organized as follows. In Section~\ref{sec:data}, we introduce the
data products used to perform our analysis and briefly describe the measurements
of the galaxy clustering and the galaxy-galaxy lensing signal. In
Section~\ref{sec:theory}, we present the theoretical background for how these
measurements can constrain cosmological parameters and the analytical halo
occupation distribution model we use to interpret the data. The results of our
main analysis and a variety of systematics tests are presented in
Section~\ref{sec:results}. We conclude in Section~\ref{sec:conc} with a summary
of our results and discuss the outlook for ongoing and future surveys.
We will assume a flat $\Lambda$CDM cosmology with $\Omega_\rmm=0.27$ when
converting redshifts to distances for performing the clustering and lensing
measurements. Throughout this paper, $\log$ denotes the $10-$based logarithm of
a quantity, the symbols $h$ and $h_{70}$ denote the Hubble constant, $H_0$
normalized by $100\kmsmpc$ and $70\kmsmpc$, respectively.

\section{Data and Measurements}
\label{sec:data}

We use the sample of galaxies compiled in Data Release 11 (DR11) of
the SDSS-III project. The SDSS-III is a spectroscopic investigation of
galaxies and quasars selected from the imaging data obtained by the
SDSS \citep{York:2000} I/II covering about $11,000$~deg$^2$
\citep{Abazajian:2009} using the dedicated 2.5-m SDSS Telescope
\citep{Gunn:2006}. The imaging employed a drift-scan mosaic CCD camera
\citep{Gunn:1998} with five photometric bands ($u, g, r, i$ and $z$)
\citep{Fukugita:1996, Smith:2002, Doi:2010}. The SDSS-III
\citep{Eisenstein:2011} BOSS project \citep{Ahn:2012, Dawson:2013}
obtained additional imaging data of about 3,000~deg$^2$
\citep{Aihara:2011}. The imaging data was processed by a series of
pipelines \citep{Lupton:2001, Pier:2003, Padmanabhan:2008} and
corrected for Galactic extinction \citep{Schlegel:1998} to obtain a
reliable photometric catalog. This catalog was used as an input to
select targets for spectroscopy \citep{Dawson:2013} for conducting the
BOSS survey \citep{Ahn:2012} with the SDSS spectrographs
\citep{Smee:2013}. Targets are assigned to tiles of diameter $3^\circ$
using an adaptive tiling algorithm designed to maximize the number of
targets that can be successfully observed \citep{Blanton:2003}. The
resulting data were processed by an automated pipeline which performs
spectral classification, redshift determination, and various parameter
measurements, e.g., the stellar mass measurements from a number of
different stellar population synthesis codes which utilize the
photometry and redshifts of the individual galaxies
\citep{Bolton:2012}. The galaxy samples in BOSS have been divided into
a low redshift LOWZ sample, and a high redshift CMASS galaxy sample.
In addition to the galaxies targetted by the BOSS project, we also use
galaxies which pass the target selection but have already been
observed as part of the SDSS-I/II project (legacy galaxies).  These
legacy galaxies are subsampled in each sector so that they obey the
same completeness as that of the CMASS sample \citep{Anderson:2014}.
In addition to the above standard reductions, we have also obtained
stellar masses for fiber collided galaxies\footnote{Galaxies which
are part of target sample but could not be allocated a fiber due to
crowding of target galaxies in dense regions.} and galaxies with
redshift failures using their own photometry but assuming that their
redshift is identical to the nearest neighbours \footnote{Nearest neighbour
corrections have been shown to accurately correct for fiber collisions above
the fiber collision scale ($\sim 0.4 \mpch$) by \citet{Guo:2012}}.

In order to define subsamples of galaxies we use stellar masses for
galaxies obtained using the Portsmouth stellar population synthesis
code \citep{Maraston:2013} with the assumptions of a passively
evolving stellar population synthesis model and a \citet{Kroupa:2001}
initial mass function. In Paper I, we divided the parent sample of
CMASS galaxies into subsamples in the stellar mass-redshift plane. The
three subsamples A, B and C that we use in our analysis all lie in the
redshift range $z\in[0.47,0.59]$ and include galaxies in the stellar
mass range $\log M_*\in[11.10,12.00]$, $\log M_*\in[11.30,12.00]$ and
$\log M_*\in[11.40,12.00]$, respectively. We will denote subsample A to be
fiducial, and test the sensitivity of our cosmological constraints to possible
incompleteness using the rest of the subsamples. The number of galaxies in
subsamples A, B and C are $400,916$, $196,578$ and $116,682$ corresponding to
number densities of $3\times10^{-4}$, $1.5\times10^{-4}$ and $0.8\times10^{-4}
~h^3\mpc^{-3}$, respectively. These numbers include galaxies that were fiber
collided and/or had failures in redshift measurements. The number density of
galaxies in each of the samples shows much less variation (less than $\sim20\%$ in
the redshift range under consideration) with redshift than the parent
sample (see Figure 1 in Paper I).

For the measurements of the galaxy-galaxy lensing signal around the
subsamples of CMASS galaxies, we must measure the tangential
distortion of background galaxies. For this purpose, we rely on the
deeper and better quality imaging data from the Canada France Hawaii
Telescope Legacy survey (CFHTLS). This information allows us to measure the
tangential distortion of background galaxies around our sample of
CMASS galaxies. In particular we make use of the photometric reduction
and image shape determinations in the publicly available CFHTLenS
catalog\footnote{\url{http://www.cfhtlens.org/astronomers/data-store}}.
The quantities needed for each galaxy, namely its
shear estimate, calibration factors, weight, and photometric redshift are
provided in the catalog \citep{Heymans:2012, Erben:2013, Miller:2013,
Hildebrandt:2012}.  Unfortunately, the overlap between the CFHTLS and the DR11
BOSS fields is limited to an area of
about 105
deg$^2$. The number
of CMASS galaxies that lie within the CFHTLS footprint is $5,084$ for our
fiducial subsample A, $2,549$ from subsample B and $1,577$ for subsample
C , respectively.

In Paper I, we presented measurements of the projected clustering of galaxies,
$w_\rmp(r_\rmp)$ for a number of different subsamples of galaxies. At fixed
redshift, we detected a clear dependence of the clustering signal on the stellar
mass of galaxies. Higher stellar mass galaxies are more clustered than lower
stellar mass galaxies. However, we also observed that the clustering of galaxies
of fixed stellar mass does not vary significantly with redshift, in particular
within the range of redshifts for the 3 subsamples considered in this paper. In
Paper I, we also measured the galaxy-galaxy lensing signal around each of our
subsamples. We also found the strength of the lensing signal to be larger for
higher stellar mass threshold samples, consistent with the expectation that
these galaxies reside in higher mass halos. In the next section we develop a
simple picture which shows how the joint measurements of clustering and lensing
of galaxies can be used to constrain cosmological parameters, as well as present
the details of the analytical model we use in order to fit a parametric model to
these measurements.

\begin{figure*} \centering{
\includegraphics[]{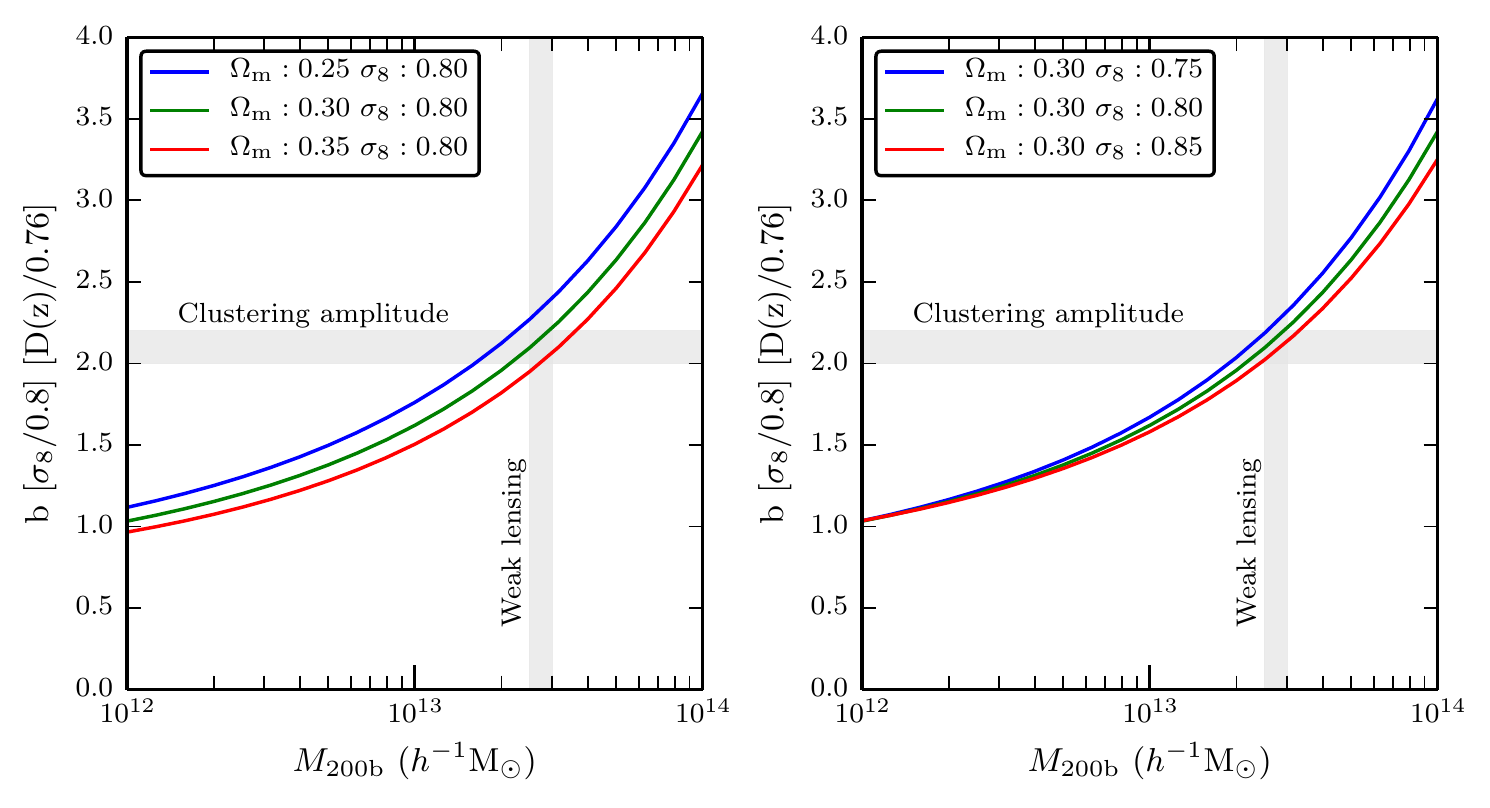}}
\caption{
    Theoretical predictions of the amplitude of the galaxy clustering signal on
    large scales as a function of the halo mass of these galaxies in different
    cosmological models at $z=0.53$. The left hand panel shows how the clustering amplitude
    varies when $\Omega_\rmm$ is increased, while the right hand panel shows the
    corresponding change when $\sigma_8$ is increased. Measurements of the
    clustering of galaxies and the galaxy-galaxy lensing signal determine the
    ordinate and the abscissa, respectively, thus allowing constraints
    on these cosmological parameters.
}
\label{fig:toy}
\end{figure*}

\begin{figure} \centering{
\includegraphics[scale=1.2]{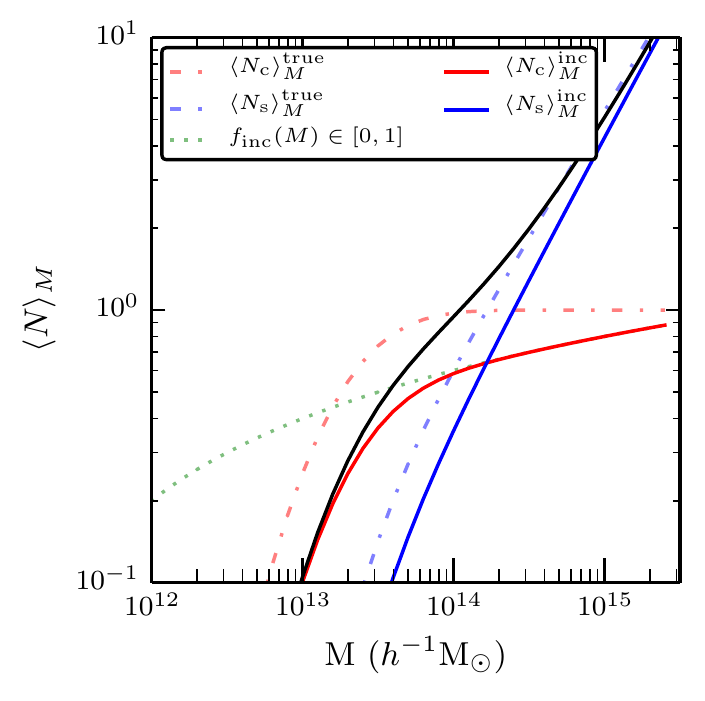}}
\caption{
An illustration of the halo occupation distribution model we use in
the analysis of this paper.  The red and blue dot-dashed lines
show the central and satellite components of the HOD appropriate for a
true stellar mass threshold sample. The green dotted line shows the
log-linear functional form we assume for parametrizing the
incompleteness in our subsample. The solid red, blue and black lines
show the HOD of centrals, satellites and all galaxies after accounting
for the incompleteness.  In total our HOD model is parametrized by 7
parameters.
}
\label{fig:nofm_toy}
\end{figure}

\section{Theory}
\label{sec:theory}

Structure formation in the concordance $\Lambda$CDM cosmological model occurs as
a result of the gravitational collapse of initial density fluctuations 
into extended halos of dark matter. The abundance of halos in different
cosmological models has a universal form when expressed as a function of peak
height, $\nu = \delta_\rmc/\sigma$, where $\delta_\rmc$ is the critical
threshold for collapse, $\sigma(M,z)$ is the variance of density
fluctuations smoothed on spatial scales corresponding to the comoving radius
from which the halo mass $M$ assembled \citep[e.g.,][]{Mo:1996,Sheth:1999}.
These halos form preferentially at the
peaks of the matter density field and hence are biased with respect to the
matter distribution. The halo bias can also be expressed as a function of the
peak height. Galaxies share the bias of the halos in which they reside.
The overall clustering amplitude at sufficiently large separation
is determined by
the products of this bias ($b$), the amplitude of the linear matter
fluctuations ($\sigma_8$) and the growth rate of fluctuations at a given
redshift ($D(z)$); $w_\rmp(r_\rmp)\propto [b\sigma_8D(z)]^2$. 

Figure~\ref{fig:toy} displays the dependence of the clustering amplitude of
halos as a function of their mass on cosmological parameters at $z=0.53$, the
average redshift of our sample.\footnote{We have used the large scale bias calibrated by
\citet{Tinker:2010} for this purpose.} In the left hand
panel, $\sigma_8$ is fixed while the matter density, $\Omega_\rmm$, is varied
for flat $\Lambda$CDM cosmological models. In the right hand panel,
$\Omega_\rmm$ is fixed while $\sigma_8$ is varied. Increasing both $\Omega_\rmm$ and
$\sigma_8$ results in a decrease of the clustering amplitude at fixed
halo mass.  The measurement of the clustering amplitude of galaxies fixes the
clustering amplitude of halos in which they reside. However, one can obtain
similar clustering amplitudes in different cosmological models by changing the
halos in which galaxies reside. This behavior is the classical degeneracy between halo
occupation distribution parameters and the cosmological parameters; one can
obtain the same clustering amplitude for galaxies by having them reside in larger
mass halos in cosmologies with larger $\Omega_\rmm$ or $\sigma_8$. The weak
lensing signal on small scales breaks this degeneracy by direct
inference of the mass of the halos, thus allowing a
determination of cosmological parameters. Given the errors in determination of
the clustering amplitude and the halo mass, we expect degeneracy in the
determination of $\Omega_\rmm$ and $\sigma_8$ such that increasing the value of
one can be compensated by decreasing the value of the other.

The above qualitative picture is valid if galaxies occupy a narrow range of
halo masses.  In reality, in our stellar mass threshold samples, galaxies
span a range in halo masses. In addition, although most galaxies are central
galaxies in their halos, some of those in our subsample are satellite galaxies.
The average clustering amplitude, $A_{\rm gal}$, is related to the bias of halos
through an integral over the halo occupation distribution of the galaxies in our
subsample,
\begin{eqnarray}
\avg{A}_{\rm gal} &=& \sigma_8 D(z) b_{\rm eff}\nonumber \\ 
&=& \sigma_8 D(z) \frac{\int \rmd M \avg{N}_M b(M,z) n(M,z)}{\int \rmd M
\avg{N}_M n(M,z)}\,, 
\label{eq:beff}
\end{eqnarray}
where the ratio of the integrals is the effective bias ($b_{\rm eff}$) of the
galaxy sample.  Similarly, the average mass of halos as determined from the
weak lensing signal needs to appropriately account for the halo occupation
distribution of the galaxies.

\subsection{Analytical HOD model}
\label{sec:hod}
We use a halo occupation distribution model \citep[hereafter
HOD;][]{Jingetal:98, PeacockSmith:00, Seljak:00, Scoccimarroetal:01,
Zheng:2005,Leauthaud:2012, vdBosch:2013, Cacciato:2013, Cacciato:2013b}, to predict the abundance, the
clustering and the lensing signal of CMASS galaxies.  We adopt an HOD model
with an explicit split of the halo occupation into central and satellite
galaxies (see Figure~\ref{fig:nofm_toy}),
\begin{equation}
\avg{N}_M =
 \avg{N_{\rm c}}_M+ \avg{N_{\rm s}}_M
\end{equation}
The mean halo occupation distribution for central galaxies is given by
\begin{equation}
\avg{N_{\rm c}}_M = f_{\rm inc}(M)\frac{1}{2}\left[1+{\rm erf}\left(\frac{\log M - \log M_{\rm
 min}}{\sigma_{\log M}}\right)\right],
\end{equation}
and that for satellite galaxies is given by
\begin{equation}
\avg{N_{\rm s}}_M = \avg{N_{\rm c}}_M\left[\frac{M-\kappa M_{\rm
min}}{M_1}\right]^\alpha
\end{equation}
when $M>\kappa M_{\rm min}$ and zero otherwise \citep[see
e.g.,][]{Zheng:2005,White:2011}. The function $f_{\rm inc}(M)$ accounts for
potential incompleteness in the selection of CMASS galaxies at the low stellar
mass end \citep[see e.g.,][]{More:2011,Reddick:2013} when compared
to a true stellar mass threshold sample. We assume a log-linear functional form
for the incompleteness function such that
\begin{eqnarray}
f_{\rm inc}(M)&=&{\rm max}[0,{\rm min}[1,1+\alpha_{\rm inc}(\log M - \log M_{\rm
inc})]]
\end{eqnarray}
This model explicitly assumes that the CMASS selection 
selects a random fraction of the stellar mass threshold galaxies,
given by $f_{\rm inc}$, from host halos at every mass scale,
equivalently, it assumes that with the CMASS color and magnitude cuts,
the selection probability for galaxies at a given stellar mass do not
depend on the environment or other properties.  

We follow the analytical framework developed in \citet{vdBosch:2013} (with a
minor extension to account for the miscentering of central galaxies with respect
to their halo centers), to predict the galaxy-galaxy clustering and the
galaxy-galaxy lensing signal, using the halo occupation distribution described
above. We briefly present the key expressions below for completeness. 

The galaxy-galaxy power spectrum, $P_{\rm gg}(k,z)$, is the Fourier transform of
the galaxy correlation function, $\xi(r,z)$ and can be expressed as a sum of
the following one- and two-halo terms,
\begin{eqnarray}
P_{\rm gg}(k,z) &=& 2\,P^{\rm 1h}_{\rm cs}(k,z) + P^{\rm 1h}_{\rm
ss}(k,z) \nonumber \\ && + P^{\rm 2h}_{\rm cc}(k,z) + 2\,P^{\rm
2h}_{\rm cs}(k,z) + P^{\rm 2h}_{\rm ss}(k,z)\,.
\end{eqnarray}
Here the subscripts ``${\rm c}$'' and ``${\rm s}$'' stand for central and satellite
galaxy, respectively. Each of these terms can be expressed in the following compact form
\begin{equation}\label{P1h}
P^{\rm 1h}_{\rm xy}(k,z) = \int\rmd M\, 
\calH_\rmx(k,M,z) \, \calH_\rmy(k,M,z) \, 
n(M,z),
\end{equation}
\begin{eqnarray}\label{P2h}
\lefteqn{P^{\rm 2h}_{\rmx\rmy}(k,z) =
\int \rmd M_1 \, \calH_\rmx(k,M_1,z) \, n(M_1,z) } \nonumber \\
& & \times \int \rmd M_2 \, \calH_\rmy(k,M_2,z) \, n(M_2,z) \,
Q(k|M_1,M_2,z)\,,
\end{eqnarray}
where `x' and `y' are either `c' (for central) or `s' (for satellite), $n(M,z)$
describes the halo mass function at redshift $z$, $Q(k|M_1,M_2,z)$ describes
the power-spectrum of haloes of masses $M_1$ and $M_2$ and accounts for the
radial dependence of bias, non-linearities in the matter power spectrum and
halo exclusion, ingredients that can be calibrated by numerical simulations
\citep[see, e.g.,][]{vdBosch:2013}. Furthermore, we have defined
\begin{equation}\label{calHc}
\calH_\rmc(k,M,z) = 
\frac{\avg{N_\rmc}_M}{\bar{n}_{\rmg}(z)} \, \left(1-p_{\rm
off}+ p_{\rm off} \exp\left[ -\frac{1}{2} k^2 (r_\rms{\cal R}_{\rm off})^2
\right] \right) \,,
\end{equation}
and
\begin{equation}\label{calHs}
\calH_\rms(k,M,z) = \frac{\avg{N_\rms}_M}{ \bar{n}_{\rmg}(z)} \,  
\tilde{u}_\rms(k|M,z)\,.
\end{equation}
Here, we have assumed that there is a fraction $p_{\rm off}$ of central galaxies
that are offset from the center of their halos \citep[see
e.g.,][]{Skibba:2011} and that the
normalized radial profile of the off-centered galaxies, with respect to the
true halo center, is a Gaussian with width relative to the scale radius,
$r_\rms$, of the halo in which they reside,
\begin{equation}
u_{\rm off}(r|M) = \frac{1}{(2\pi)^{3/2} (r_\rms{\cal R}_{\rm off})^3}\exp\left[
-\frac{1}{2}\left(\frac{r}{r_\rms{\cal R}_{\rm off}}\right)^2 \right]\,.
\end{equation}
The Fourier transform of $u_{\rm off}(r|M)$ is $\exp[-k^2(r_s{\cal R}_{\rm
off})^2/2]$, and the
quantity $\tilde{u}_\rms(k|M)$ in $\calH_\rms(k,M,z)$ is the Fourier transform
of an Navarro-Frenk-White \citep[][hereafter NFW]{Navarro:1996} profile for a
halo of mass $M$ \citep[see also][for a similar model]{Hikage:2013}.  We also
assume that the normalized number density
profile of satellite galaxies follows the NFW profile\footnote{We have examined
models which allow the satellite galaxies to have a concentration which is
different from that of the dark matter distribution. The clustering signal on
small scales is sensitive to this parameter. We model the clustering signal on
scales larger than $r_\rmp>0.85\mpch$ where the impact of this parameter is
minimal. We have verified that the cosmological constraints are robust to the
inclusion or exclusion of such a parameter.}. The number density of galaxies,
$\bar{n}_{\rmg}(z)$, is given by
\begin{equation}
\bar{n}_{\rmg}(z)=\int \avg{N}_M n(M,z) \rmd M
\end{equation}
We will assume a $\sim$20 percent fractional error on the abundances, since
our stellar mass cuts yield a roughly constant abundance with redshift,
with $\sim 20$ percent level fluctuations.  
In the presence of parameters to model the incompleteness we do
not expect the abundances to influence the cosmological constraints in a
significant manner.

The real-space correlation function, $\xi_{\rm gg}(r,z)$, can be obtained by an
inverse Fourier transform of the galaxy-galaxy power spectrum \footnote{We
integrate over all $k$ while carrying out the Fourier transform, but assume a
single redshift for the calculation.}. We use a
modified version of the large scale redshift space distortion model presented by
\citet{Kaiser:1987} to predict $\xi^z_{\rm gg}(r_\rmp,\pi,z)$ from $\xi_{\rm
gg}(r,z)$ \citep[see][for details]{vdBosch:2013}, and integrate along the
line-of-sight,
\begin{equation}
w_\rmp(r_\rmp)=2 \int_{0}^{\pi_{\rm max}} \xi(r_\rmp,\pi) \,\rmd \pi\,.
\label{eq:wp}
\end{equation}
to calculate the projected correlation function. This modified
model accounts for residual redshift space distortions on large scales
due to finite value of $\pi_{\rm max}$ \citep[see
e.g.,][]{Norberg:2009, Baldauf:2010, More:2011b, vdBosch:2013}. The upper limit
for the line-of-sight integration we adopt is $\pi_{\rm max}=100\mpch$, thus
mimicking the integration limit adopted in the measurements in Paper I for the
subsamples of galaxies we use. 

The galaxy-galaxy lensing signal is a probe of the excess surface density,
\begin{equation}
\Delta \Sigma(r_\rmp)=\avg{\Sigma(<r_\rmp)}-\bar{\Sigma}(r_\rmp)\,
\label{eq:dSigma}
\end{equation}
The surface density $\Sigma(r_\rmp,z)$
can be obtained by projecting the galaxy-matter correlation function,
$\xi_{gm}(r,z)$, using
\begin{equation}
\Sigma(r_\rmp,z) = \int_{R}^{\infty} \bar{\rho}\,[1+\xi_{\rm gm}(r,z)]
\frac{2\,r\,\rmd r}{\sqrt{r^2-r_\rmp^2}}\,.
\label{eq:sigproj}
\end{equation}
In order to predict the galaxy-matter cross power spectrum, we adopt the HOD
model framework. The cross power spectrum is given by the sum of the following
one- and two-halo terms
\begin{equation}\label{Pgm}
P_{\rm gm}(k,z) = P^{\rm 1h}_{\rm cm}(k,z) + P^{\rm 1h}_{\rm sm}(k,z) 
+ P^{\rm 2h}_{\rm cm}(k,z) + P^{\rm 2h}_{\rm sm}(k,z)\,.
\end{equation} 
Each of the above terms can be calculated using Eqs.~(\ref{P1h})-(\ref{P2h}),
where `x' is `m' (for matter) and `y' is either `c' (for central) or `s' (for
satellite). For the matter component, we define
\begin{equation}\label{calHm}
\calH_\rmm(k,M,z) = {M \over \bar{\rho}_{\rmm}(z)} \,
\tilde{u}_\rmh(k|M,z)\,,
\end{equation}
where $\tilde{u}_\rmh(k|M,z)$ is the Fourier transform of the normalized
density distribution of matter within a halo of mass $M$, and
$\bar{\rho}_{\rmm}(z)$ denotes the average comoving density of the
Universe at redshift $z$. The galaxy-matter correlation function can be
obtained by an inverse Fourier transform of the galaxy-matter power
spectrum.

The lensing signal is sensitive to the total matter content of galaxies
including both the dark matter and the baryonic components.  Therefore, we will
also consider the matter component in the lens galaxy, which includes stars and
gas\footnote{The gas fractions around high stellar mass galaxies are expected
to be small, so we assume all the baryonic mass is in stars.}. At distances
close to the lensing galaxy, these terms could dominate the lensing signal. We
assume that the effect of the matter component in the lens galaxy can be
considered as a point mass contribution located at the position of the lens
galaxy,
\begin{equation}
\Delta \Sigma_\ast(r_\rmp) = \frac{\tilde M_\ast}{\pi r_\rmp^2}\,,
\end{equation}
where $\tilde M_\ast$ is in units of $h^{-1}M_\odot$ and the projected radius
$r_\rmp$ is in units of $h^{-1}\pc$. The stellar population synthesis models
(hereafter SPS) infer the stellar mass, $M_\ast$ based upon the luminosity and
mass-to-light ratio of stars. These masses therefore have the units of
$h^{-2}M_\odot$, and this stellar mass is related to the baryonic lensing mass
by $\tilde M_\ast=M_\ast/h$. Our models adopt the parameter $M_\ast$ to allow a
comparison of this mass with the stellar mass measurements from the SPS
models.

In addition to the HOD parameters, our analytical predictions also depend upon
the cosmological parameters, via the halo mass function, the halo bias
function, and the cosmology dependence of the concentration-mass relation
\citep{vdBosch:2013}. There are a number of simulation-calibrated ingredients
required to use the analytical expressions in this section. For
the sake of completeness and reproducibility, we list each of them below. We
assume the halo masses to be $200$ times overdense with respect to the
background matter density. We use the halo mass function calibration of 
\citet{Tinker:2008} and large scale bias calibration of \citet{Tinker:2010} for this particular definition. The radial dependence of halo bias was calibrated by
\citet{Tinker:2005} for friends-of-friends halos. We use an appropriate
modification to take into account the spherical overdensity definition of halos
and the effects of halo exclusion \citep[see][]{vdBosch:2013} which allows us to
calculate $Q(k|M_1,M_2,z)$. We use a nuisance parameter $\psi$ to marginalize
over the uncertain description of radial dependence of halo bias. This parameter
governs the behaviour of the prediction in the transition regime between
one- and two-halo terms \citep[see][for details]{vdBosch:2013}.  The
concentration of dark matter halos is assumed to follow the concentration-mass
relation calibration presented by \citet{Maccio:2008}. We allow for a
normalization parameter $R_\rmc$ which characterizes deviations from this
fiducial relation and assign it a prior of $1.00\pm0.20$. 
We also assume that
the number density profile of satellite galaxies will share the same
concentration as that of the dark matter distribution.
We have checked that
increasing the width of the prior on the $R_\rmc$ to $0.30$, or allowing the
concentration parameter for satellite galaxy distribution to differ from the
dark matter distribution does not significantly affect our
results. 
In addition, we also
allow for a $2.5$ percent uncertainty in the modeling of the projected
clustering signal (the overall amplitude) to account for the inaccuracies of the
model. We do not explicitly include a corresponding systematic uncertainty in
the lensing calibration, since the statistical errors on that measurement are
already sufficiently large that they are the dominant source of error, compared
to the statistical error on the systematic shear calibration
correction \citep{Miller:2013}.

\subsection{Cosmological parameter dependence of the measurements}
\label{sec:cosdep}
\begin{figure*} \centering{
\includegraphics[]{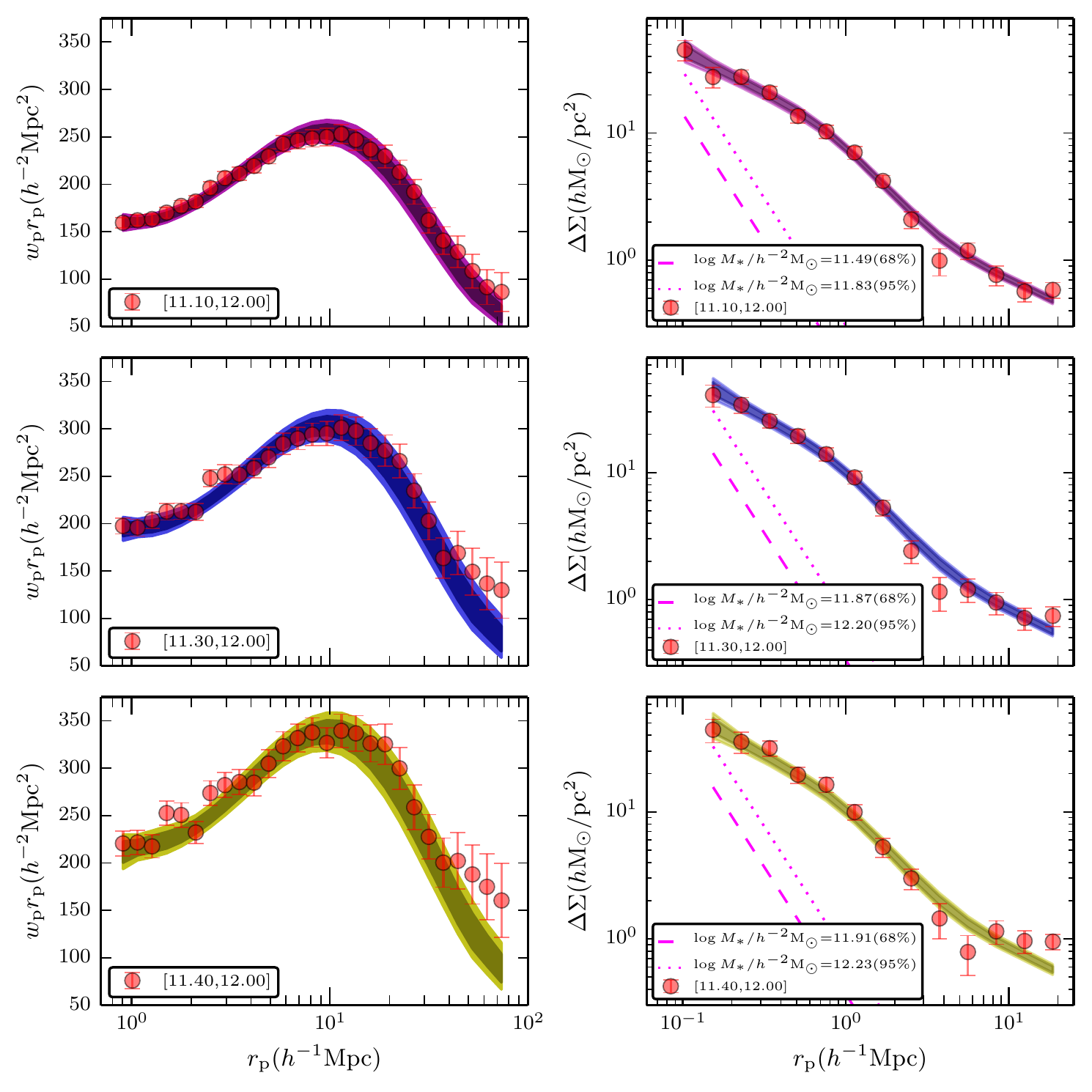}}
\caption{
    The 68 and 95 percent confidence intervals of the fits to the clustering
    measurements (left panels) and the lensing measurements (right panels)
    obtained from the HOD modeling exercise for subsamples A, B and C are shown
    in the top, middle and bottom rows, respectively. The intervals are
    obtained by projecting uncertainties in all the model parameters
    around the best-fit model. The reduced $\chi^2$ for the best fit models
    in the three cases are $0.8$, $1.3$ and $1.5$ for 31, 31 and 30 degrees
    of freedom, respectively (see text for details). The 68 and 95 percent upper
    limits on the stellar mass of galaxies are also shown in the right hand
    panels.
}
\label{fig:wp_esd_fit}
\end{figure*}

\begin{figure*} \centering{
\includegraphics[width=1.0\textwidth]{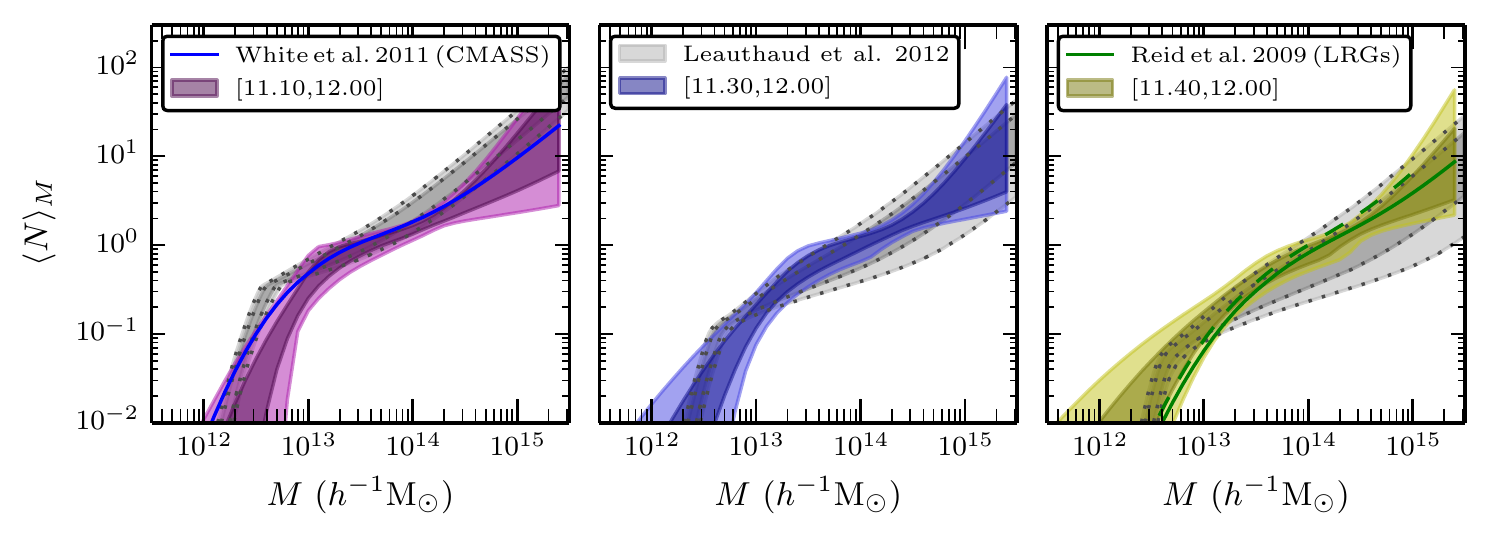}}
\caption{
The 68 and 95 percent confidence intervals of the halo occupation distribution
of CMASS galaxies in samples A, B and C obtained from our modeling exercise are
shown in the three different panels, respectively. The HOD constraints displayed
here are marginalized over the uncertainty in the cosmological parameters.
Subsamples A, B and C occupy progressively more massive halos. The results for
subsample A can be contrasted with results from \citet{White:2011} who constrain
the HOD using the clustering of an early data release of CMASS galaxies (shown
as solid blue line in left hand panel). The halo occupation distribution for
sample C is consistent with the HOD obtained by \citet[][shown as green solid
line in the right hand panel]{Reid:2009}, based on a counts-in-cylinder analysis
of the LRG sample of galaxies. The green dashed line in the right hand panel
shows the result of a simple attempt to correct for the differences in the mean
redshift of LRGs and CMASS galaxies by adjusting the masses of LRGs at $z=0.3$
to the masses of their progenitors at $z=0.53$. The gray shaded bands in each of
the subsample show the constraints on the HOD obtained by \citet{Leauthaud:2012}
from COSMOS data, employing the same stellar mass cuts in the galaxy selection.
The similarity of our HOD constraints, especially for the higher stellar mass
threshold samples, implies that the magnitude of potential incompleteness
effects in our analysis decrease with increasing stellar mass threshold.
}
\label{fig:nofm}
\end{figure*}

The abundance, clustering and the lensing measurements depend upon
the fiducial cosmological model, ${\cal C}^{\rm fid}$, that we have assumed to
convert the angular and redshift differences in the positions of galaxies to
their comoving separation. We follow \citet{More:2013a}, in order to account for
this dependence. For a given cosmological model ${\cal C}$, we
multiply the predicted abundance of galaxies $\bar{n}_\rmg({\cal C})$ in the
redshift bin $[z_2,z_1]$ by the ratio of the comoving volumes
\begin{equation}
\bar{n}_\rmg' = \bar{n}_\rmg({\cal C})\frac{\chi^3(z_2,{\cal
C})-\chi^3(z_1,{\cal C})}{ \chi^3(z_2,{\cal
C}^{\rm fid})-\chi^3(z_1,{\cal C}^{\rm fid})}
\end{equation}
in order to compare it to abundance measured assuming ${\cal C}^{\rm fid}$.
Here, $\chi(z)$ denotes the comoving distance to redshift $z$.  For the
clustering and the lensing signals, we first calculate $w_\rmp$ and $\Delta
\Sigma$ at comoving separations $r_\rmp'$, which are related to the projected
comoving separation $r_\rmp^{\rm fid}$ at which the measurements were performed
by
\begin{equation}
r_\rmp'=r_\rmp^{\rm fid} \left[\frac{\chi(\bar z,{\cal C})}{\chi(\bar z,{\cal C}^{\rm
fid})}\right]
\end{equation}
where $\chi^{\rm fid}$ and $\chi$ denote the comoving distance to the median
redshift $\bar z$ in ${\cal C}^{fid}$ and ${\cal C}$, respectively. This
calculation accounts for the difference in the conversion of angular differences
between galaxies to comoving separations. We need to further change the
amplitude of the predictions of both $w_\rmp$ and $\Delta\Sigma$, such that
\begin{eqnarray}
\tilde{w}_\rmp(r_\rmp) = w_\rmp(r_\rmp') \left[\frac{E(\bar z)}{E^{\rm fid}(\bar z)}\right]  \\
\tilde{\Delta \Sigma}(r_\rmp)= \Delta \Sigma(r_\rmp') \left[\frac{\Sigma_{\rm crit}(\bar
z,z_\rms)}{\Sigma_{\rm crit}^{\rm fid}(\bar z,z_\rms)}\right],
\end{eqnarray}
where the former equation accounts for the cosmology dependence of the
conversion from redshift difference to comoving line-of-sight distances, while
the latter corrects for the cosmology dependence of $\Sigma_{\rm crit}$, which is
used to calculate the lensing signal from the measured ellipticities. We use a
fixed source redshift to calculate $\Sigma_{\rm crit}(\bar z,z_\rms)$ due to its
weak dependence on the choice of the source redshift. We compare these modified
predictions with the measurements. The validity of this procedure to capture the
cosmological dependence of the measurements procedure was verified in
\citet{More:2013a}.

\begin{figure*}
\centering{
\includegraphics[]{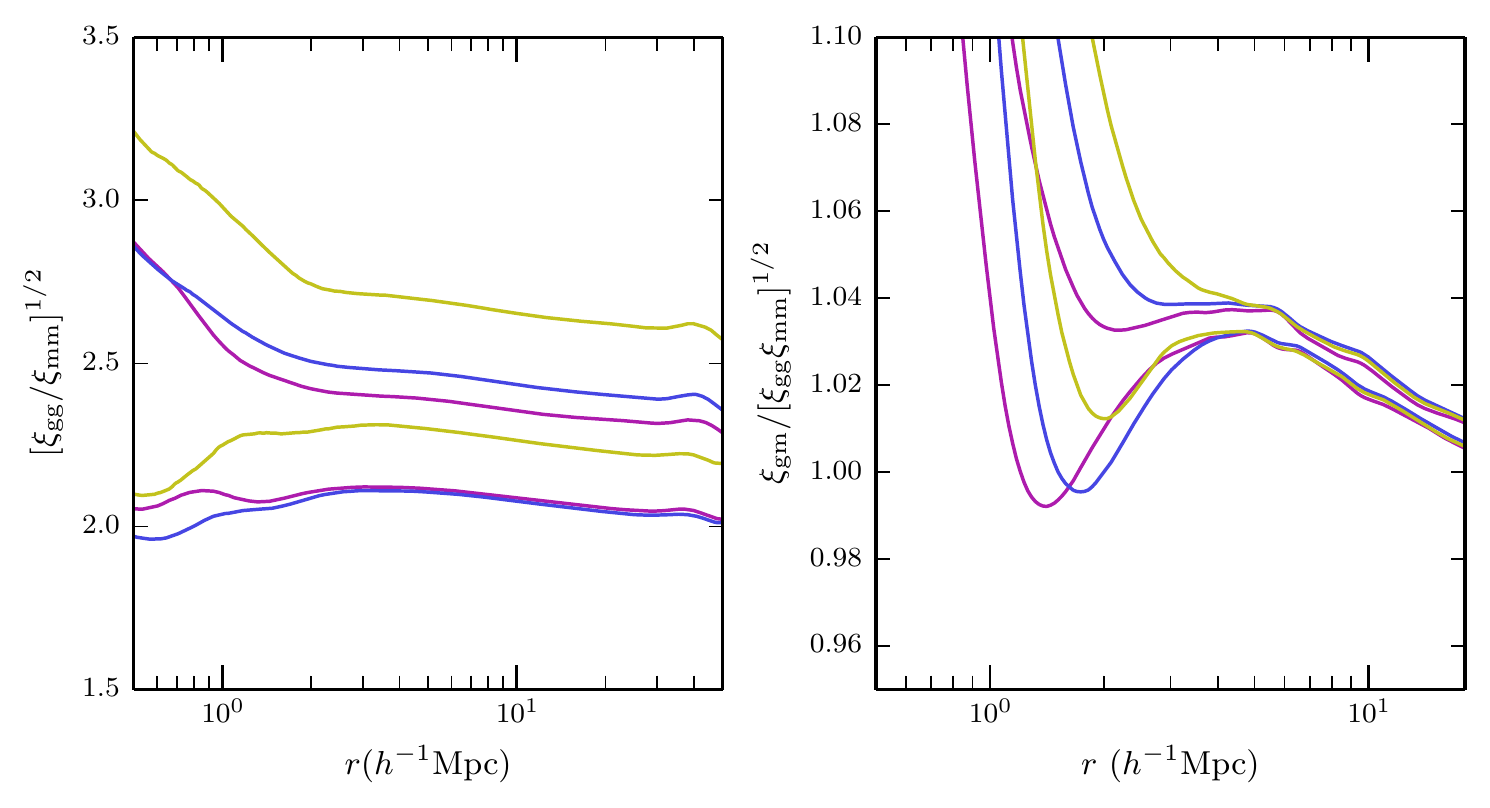}
\caption{
    The 68 percent confidence limits on the scale-dependence of bias (left
    panel) and the cross-correlation coefficient (right panel) of our subsamples
    A, B and C are enclosed by the magenta, blue and yellow lines, respectively.
    The bias tends to a constant value on large scales, and the
    cross-correlation coefficient shows small but significant deviations from
    unity at small scales.
\label{fig:bias_crossr}
}
}
\end{figure*}

\begin{figure*} \centering{
\includegraphics[]{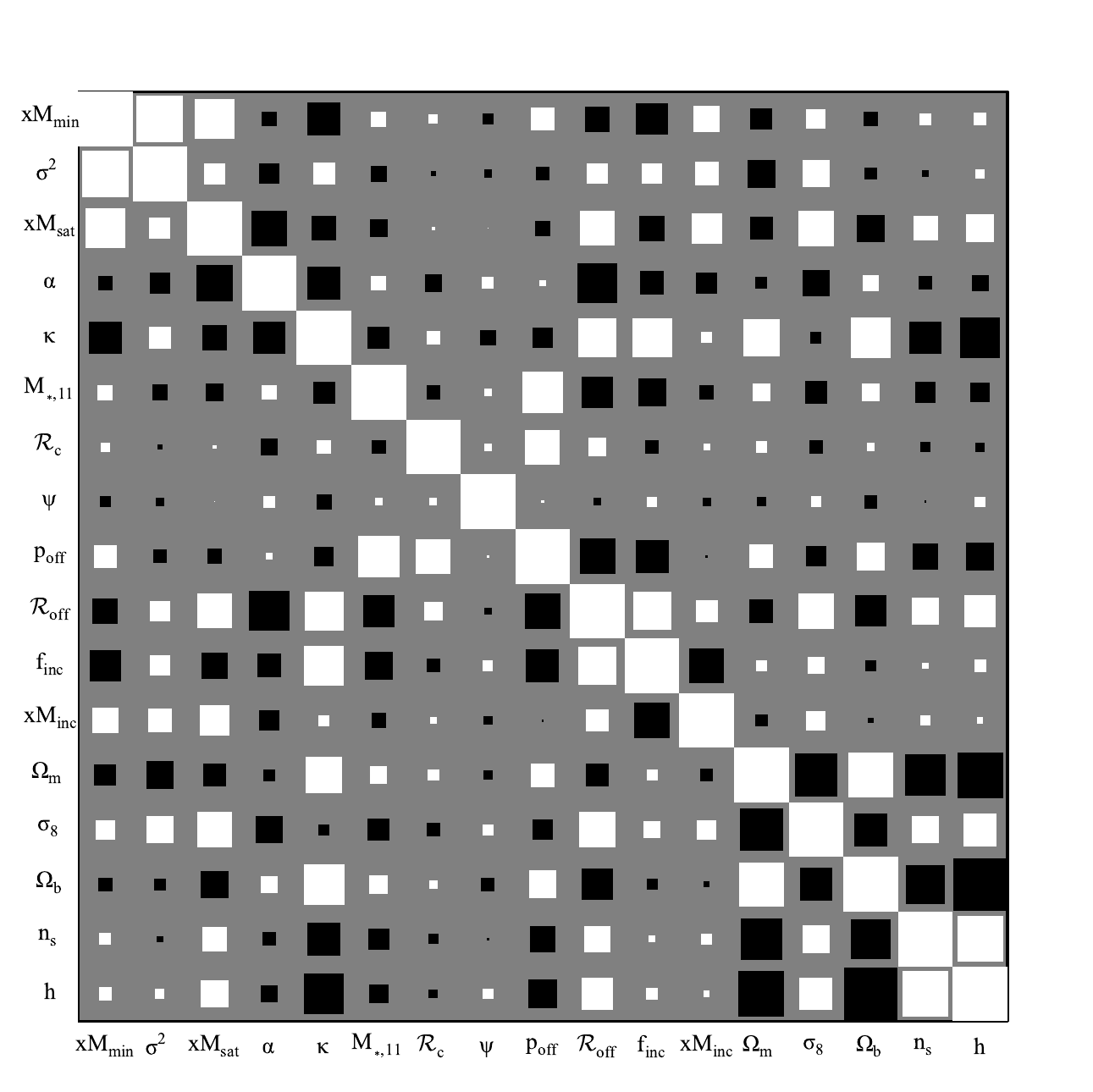}}
\caption{
    The degeneracies between different parameters constrained by our modeling
    of the clustering and lensing of CMASS galaxies with stellar mass threshold
    $\log M_*>11.10$. Positive correlations in the posterior are displayed with
    white squares and the negative correlations are displayed with black
    squares. The size of the squares indicates the strength of correlation.
}
\label{fig:Hinton_11_10}
\end{figure*}

\begin{figure*} \centering{
\includegraphics[]{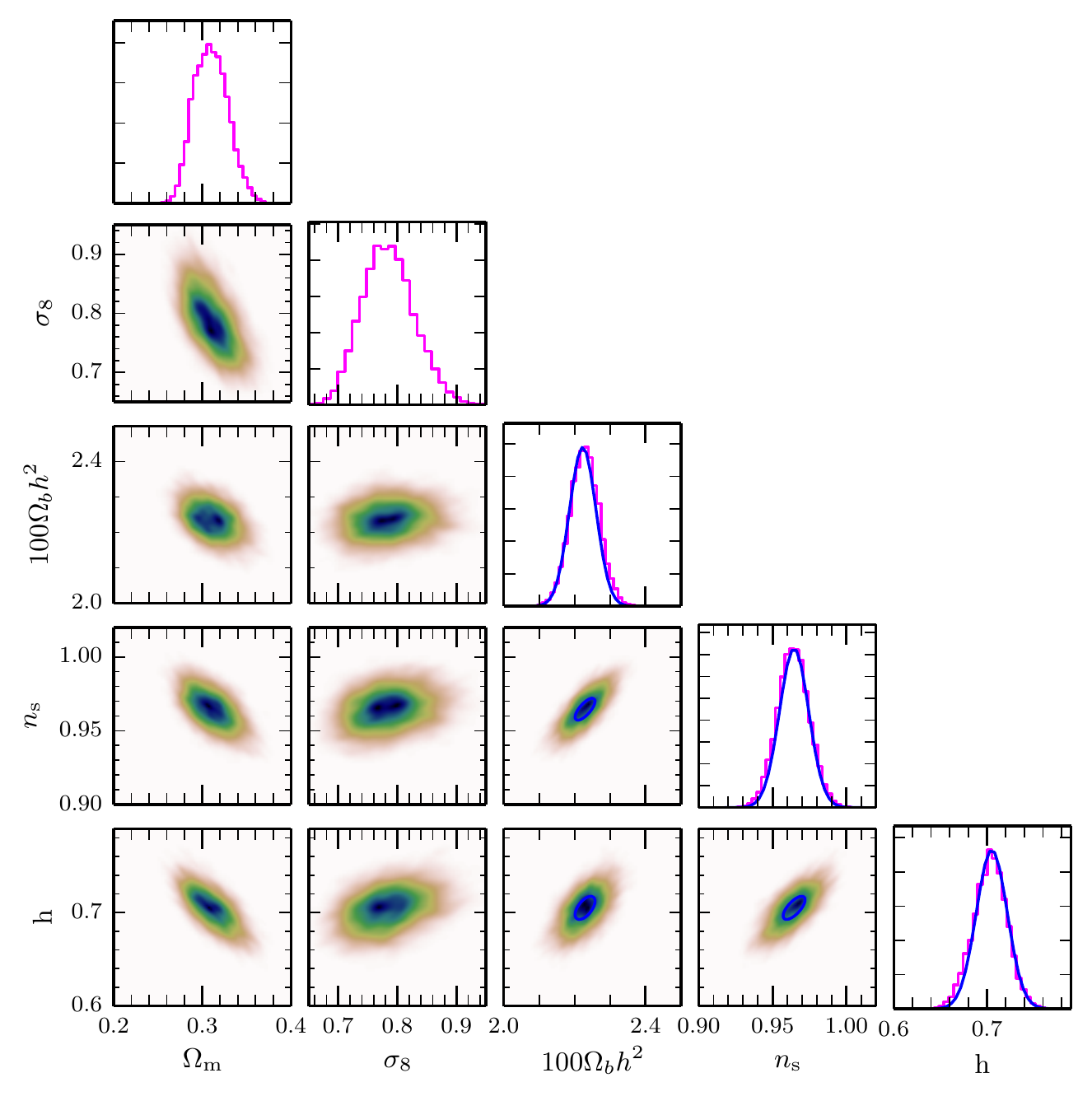}}
\caption{
    The degeneracies between different cosmological parameters as inferred from
    the analysis of the clustering and lensing signal from our fiducial
    subsample A are shown with shaded contours in the off-diagonal panels. The
    magenta histograms in the diagonal columns represent the posterior
    distribution of each of our cosmological parameters. The blue distributions
    in the diagonal panels and the blue contour (68 percent confidence) in some
    of the off-diagonal panels represent the prior information we adopt on the
    parameters $[\Omega_\rmb h^2,n_\rms, h]$.
}
\label{fig:cosmo_degen}
\end{figure*}

\begin{figure} \centering{
\includegraphics[]{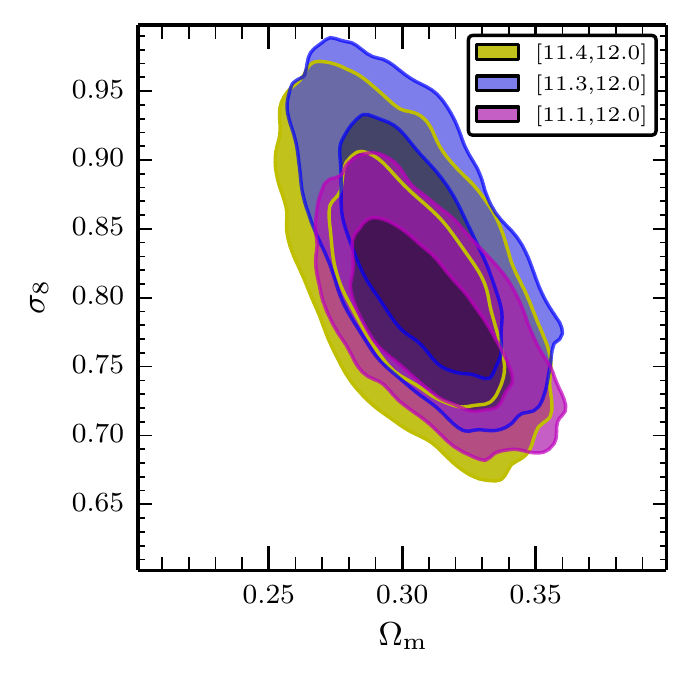}}
\caption{
    The constraints in the $\Omega_\rmm-\sigma_8$ plane obtained from the
    analysis of the different stellar mass threshold samples are shown using
    contours of different colours. As the higher stellar mass subsamples are
    expected to be more complete, this provides a systematic check of the
    possible bias in our cosmological constraints due to incompleteness.
}
\label{fig:Omm_s8_systematics}
\end{figure}

\begin{figure} \centering{
\includegraphics[]{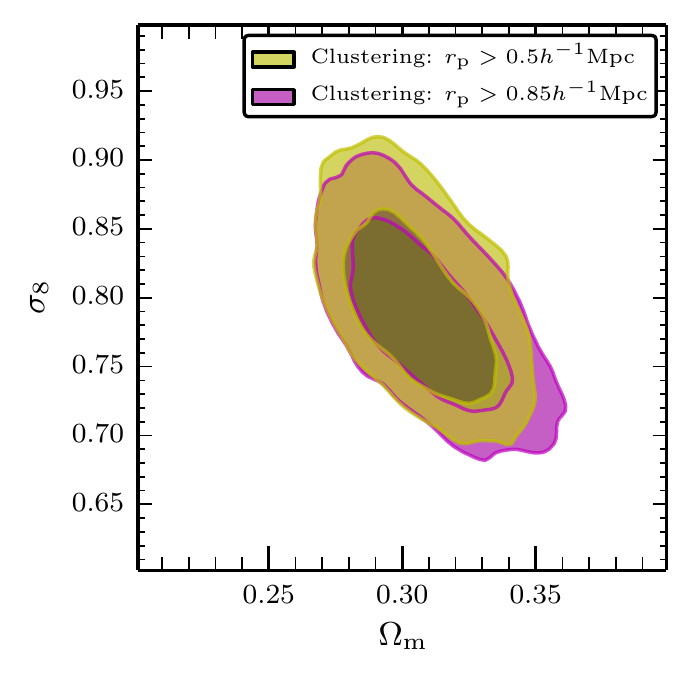}}
\caption{
    The constraints in the $\Omega_\rmm-\sigma_8$ plane obtained from the
    fiducial subsample A, but restricting the projected clustering signal to
    different scales. The magenta contours correspond to our fiducial analysis
    with $r_\rmp>0.85\mpch$, while the yellow contours correspond to $r_\rmp>0.5
    \mpch$. Inclusion of the small scale clustering information does not
    cause a drastic improvement in the cosmological constraints.
}
\label{fig:rp_systematics}
\end{figure}

\begin{figure} \centering{
\includegraphics[]{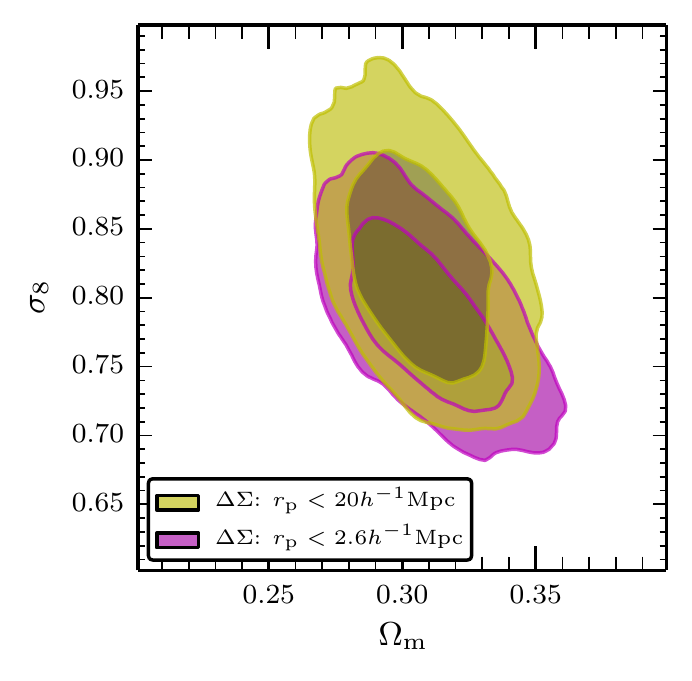}}
\caption{
    The constraints in the $\Omega_\rmm-\sigma_8$ plane obtained from the
    fiducial subsample A, but restricting the galaxy-galaxy lensing signal to
    small scales. The magenta contours correspond to our fiducial analysis
    with $r_\rmp<20.0\mpch$, while the yellow contours correspond to
    $r_\rmp<2.6 \mpch$. By excluding the large scales, we are less sensitive to
    the two halo-term in the lensing signal, which could be affected if the
    CFHTLS sample is not representative of our CMASS subsample.
}
\label{fig:esdmax}
\end{figure}

\begin{figure*} \centering{
\includegraphics[]{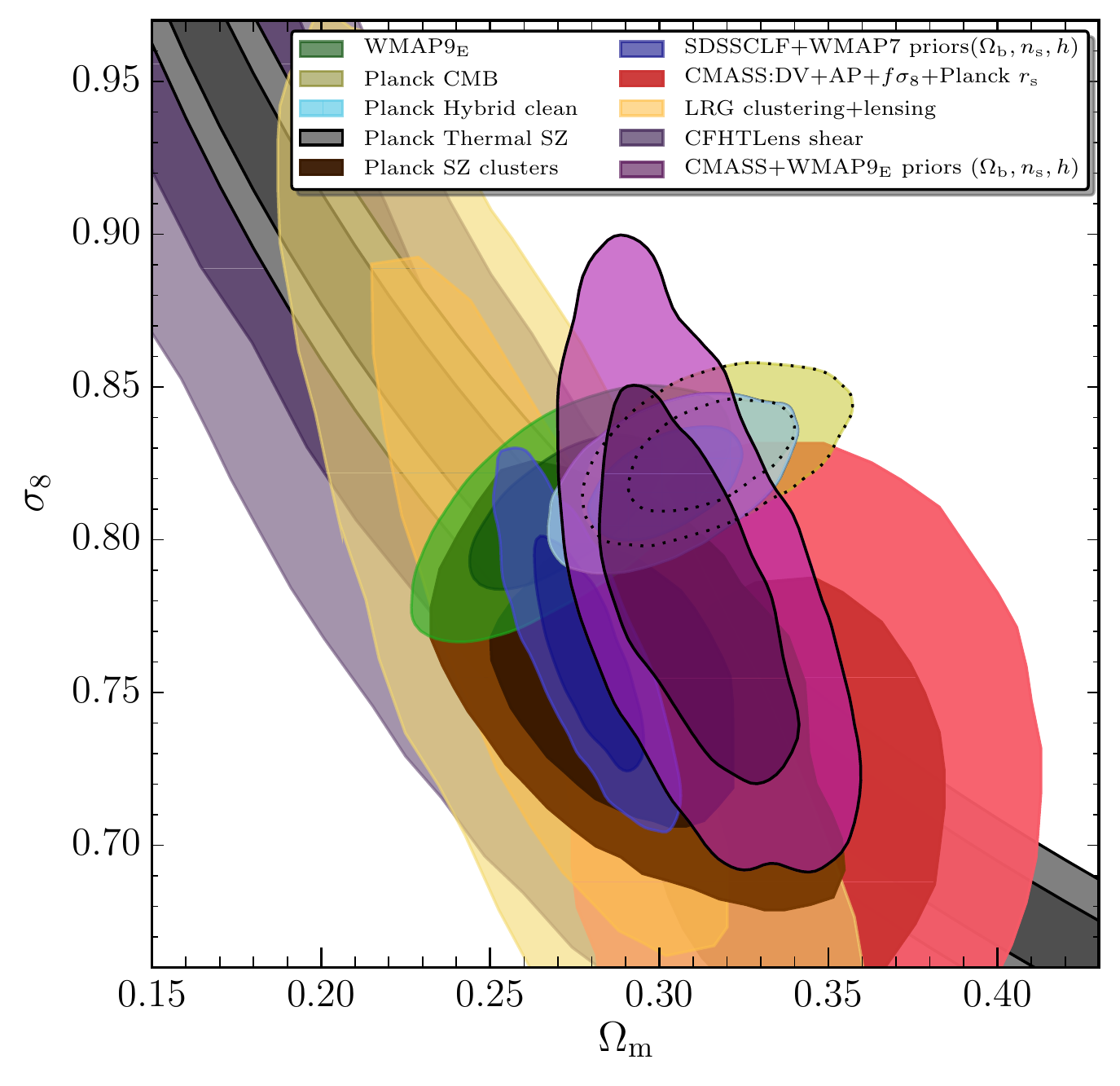}}
\caption{
The 68 and 95 percent confidence constraints on the matter density parameter,
$\Omega_\rmm$ and the fluctuation amplitude parameter, $\sigma_8$  obtained from
the analysis of the clustering and lensing measurements of our fiducial sample
are shown using magenta contours. These can be compared with results from other
cosmological probes such as the analyses of the cosmic microwave background
carried out by the WMAP team (green shaded regions, denoted as
WMAP$9_{\mbox{E}}$), the Planck team (chrome
yellow shaded regions with dotted contours; Planck CMB) and a reanalysis of Planck data by
\citet[][light blue shaded regions; Planck Hybrid
 clean]{Spergel:2013}, the SZ cluster abundances (brown shaded regions;
 Planck SZ clusters) and the
 thermal SZ power spectrum (gray shaded regions; Planck Thermal SZ)
 carried out by the Planck team,
the joint analysis of clustering and lensing of the SDSS main galaxy sample
(dark blue; SDSSCLF+WMAP7) by \citet{Cacciato:2013}, and that of the LRG sample (yellow;
LRG clustering+lensing) by \citep{Mandelbaum:2013}, the joint analysis of redshift
space distortions , BAOs and the Alcock Paczynski test using CMASS galaxies (red
 shaded regions; CMASS:DV+AP+$f\sigma_8$+Planck $r_s$) by \citep{Beutler:2013}
 and the tomographic weak lensing signal (violet shaded
 regions; CFHTLens shear) by \citep{Heymans:2013}.
 The figure shows that our results are consistent, complementary and
competitive with constraints from different cosmological probes.
}
\label{fig:Omm_s8}
\end{figure*}

\begin{figure*} \centering{
\includegraphics[width=14.0cm]{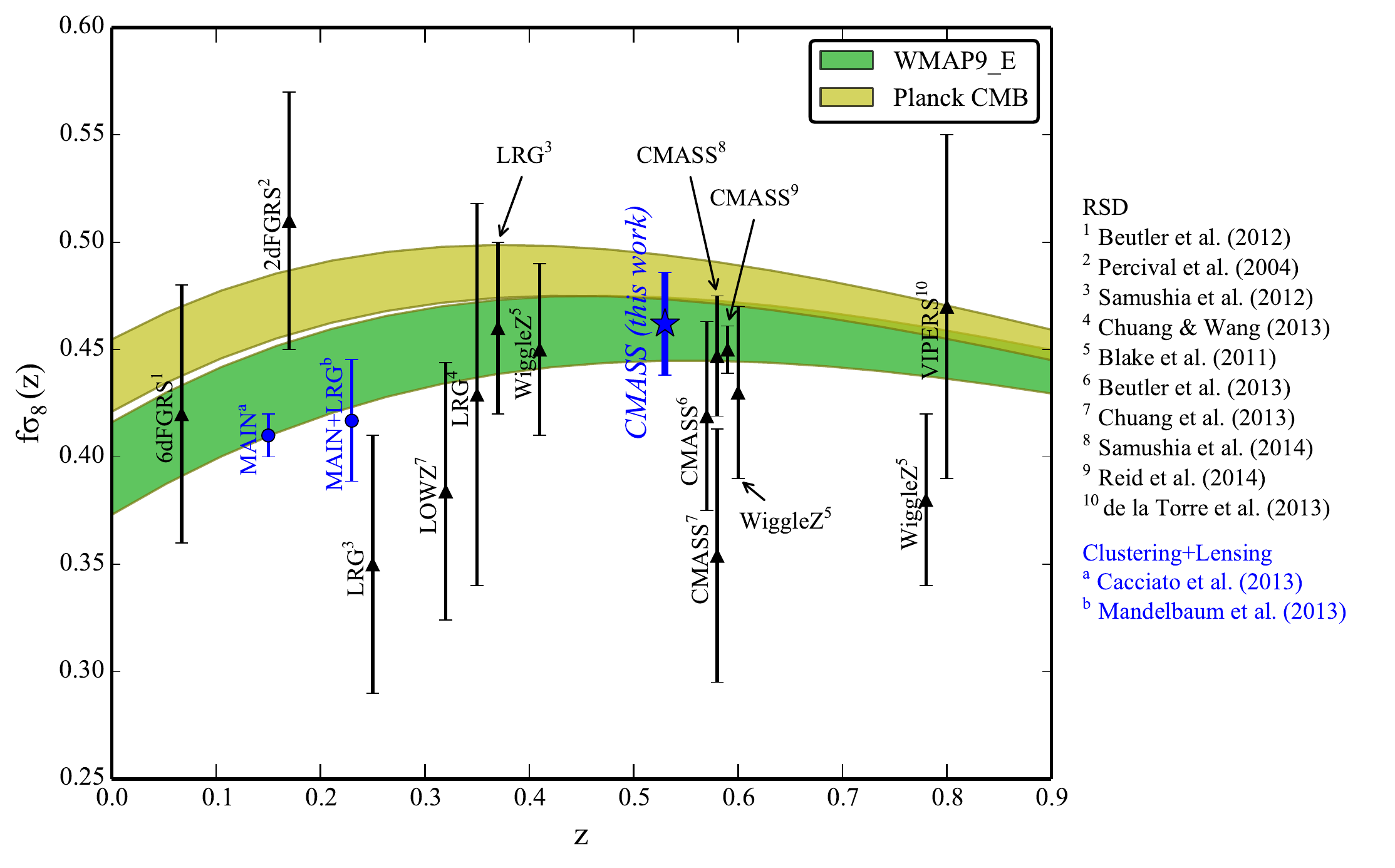}} \caption{Blue dots denote
$f\sigma_8(z)$ constraints calculated from our measurement and other
joint analyses of clustering and lensing measurements. Black triangles
denote $f\sigma_8(z)$ constraints based on various RSD measurements. Some of
the RSD CMASS measurements are shifted along the redshift-axis for clarity
($+0.01$ for references 6 and 8 and $+0.02$ for reference 9). Error bars show
68 percent confidence level. Green (Yellow) shaded region shows $f\sigma_8$
predictions with 68 percent confidence intervals based on the WMAP9 (Planck) CMB
measurement. Our result is the highest redshift cosmological analysis using a
joint analysis of the clustering and lensing signal of galaxies.}  
\label{fig:fsigma8}
\end{figure*}

\subsection{Summary of model parameters}

The analytical model we use to describe the measurements has 17
parameters. The first set of 5 parameters, $M_{\rm min}, \sigma^2,
M_{\rm sat}, \alpha, \kappa$, describes the halo occupation distribution of
galaxies. The parameter $M_{*,11}$ describes the average stellar mass of
galaxies in units of $h^{-2}M_\odot$, while ${\cal R}_\rmc$ is the
normalization of the concentration mass relation with respect to the one
obtained from simulations.  We have 5 nuisance parameters: $\psi,p_{\rm
off},{\cal R}_{\rm off},\alpha_{\rm inc},M_{\rm inc}$. Finally there are 5
cosmological parameters: $\Omega_\rmm,\sigma_8,\Omega_\rmb h^2, n_\rms$ and $h$.
We let $\Omega_\rmm$ and $\sigma_8$ be completely free but use priors on the
latter three cosmological parameters from the joint likelihood of the cosmic microwave
background analysis of the WMAP 9-year data and the high resolution CMB
measurements from the South Pole Telescope (SPT) and the Atacama Cosmology
Telescope (ACT) obtained by \citet{Hinshaw:2013}. We will denote this
combination of cosmological priors as WMAP9\_E. Our samples A and B have 27 measurements of
$w_\rmp$, 14 measurements of $\Delta\Sigma$ and one of the abundance for
the subsample each, used for the analysis. We have 17 total model parameters, with priors 
on the 3 cosmological parameters $(\Omega_\rmb h^2, n_\rms, h)$ and on the 3
parameters $\alpha,{\cal R}_\rmc,\psi$. Therefore, the total number of degrees of freedom
are $27+14+1-17+6=31$. The number of degrees of freedom are 30 for subsample C
which does not include the innermost bin in the lensing measurement.
We perform a Bayesian inference of cosmological parameters given the
measurements using a Markov chain Monte Carlo (hereafter MCMC) analysis. In particular,
we use the affine invariant sampler of \citet{Goodman:2010} as implemented by
the software {\it emcee} to navigate the parameter space
\citep{ForemanMackey:2013}. 

\section{Results}
\label{sec:results}

We now present the results of our HOD analysis and the different
systematic tests we have performed. The 68 percent confidence intervals on each
of our parameters from the analysis of the abundance, clustering and the
lensing signal of the three stellar mass subsamples we use are
listed in Table~\ref{table:post}. 
The top left- and right-hand panels of Figure~\ref{fig:wp_esd_fit} show
the projected clustering data and the lensing data with errorbars for
the fiducial subsample of galaxies, while the middle and bottom panels
display the corresponding results for subsamples B and C, respectively.
The dark and light colored shaded regions
denote the 68 and 95 percent confidence intervals obtained from the MCMCs,
respectively, marginalizing over all of the model parameters.  The overall shape
of the clustering and lensing signals is reproduced well by the model.  The
clustering and lensing predictions are able to successfully reproduce the
increasing strength with stellar mass thresholds, as in the observations.  The
model is also able to reproduce the shape of the weak lensing signal and the
observed transition between the one- and two-halo regimes.  There is a hint that
the model has some difficulty reproducing the high amplitude of the clustering on
large scales for the higher stellar mass threshold samples. However, note that
the large scales have significant covariance. The reduced $\chi^2$ for
the best fit models for the fiducial subsample is $0.8$, while those of the
subsamples B and C are $1.3$ and $1.5$, respectively. The
probabilities to exceed the $\chi^2$ values by chance given the number
of degrees of freedom are equal to $76\%$, $10\%$ and $4\%$, respectively.

The right hand panels of Figure~\ref{fig:wp_esd_fit} present the 68 and 95
percent upper limits on the average stellar mass of the galaxies in each
subsample using dashed and dotted lines, respectively. The 68 percent confidence
limits from the weak lensing modeling are $\avg{M_*}<3\times10^{11} h^{-2}
M_\odot$, $ \avg{M_*}<7.4\times 10^{11} h^{-2} M_\odot$ and $\log \avg{M_*}<8.1
\times 10^{11} h^{-2} M_\odot$ for the progressively larger stellar mass
subsamples, respectively. These limits on the average stellar masses from weak
lensing have been marginalized over uncertainties in the cosmological model. We
compare our results with the estimates of stellar masses for our subsample of
galaxies from stellar population synthesis models with a variety of different
assumptions and codes in Table~\ref{table:compare}. The limits could be improved
in the near future if the weak lensing signal can be measured to even smaller
scales (Kobayashi et al., submitted). We expect such measurements to
potentially constrain stellar population synthesis models.

\renewcommand{\tabcolsep}{0.2cm}
\begin{table}
\begin{center}
\caption{Posterior distribution of parameters from 
the MCMC analysis
}
\label{table:post}
\begin{tabular}{cccc}
\hline\hline
 & & Subsample &  \\
Parameter & [11.10,12.00] & [11.30,12.0] & [11.40,12.0] \\
\hline
$\log M_{\rm min}$ & $13.13_{-0.13}^{+0.13}$ & $13.45_{-0.15}^{+0.15}$ & $13.68_{-0.16}^{+0.16}$ \\ 
$\sigma^2$ & $0.22_{-0.15}^{+0.14}$ & $0.45_{-0.25}^{+0.24}$ & $0.79_{-0.39}^{+0.41}$ \\ 
$\log M_{1}$ & $14.21_{-0.13}^{+0.11}$ & $14.51_{-0.15}^{+0.17}$ & $14.56_{-0.22}^{+0.27}$ \\ 
$\alpha$ & $1.13_{-0.45}^{+0.38}$ & $1.14_{-0.51}^{+0.47}$ & $1.00_{-0.43}^{+0.45}$ \\ 
$\kappa$ & $1.25_{-0.69}^{+0.71}$ & $0.85_{-0.67}^{+0.67}$ & $1.19_{-0.96}^{+0.99}$ \\ 
$M_{*,11}$ & $0.0^{+3.1}$ & $0.0^{+7.3}$ & $0.0^{+8.1}$ \\ 
${\cal R}_\rmc$ & $0.98_{-0.20}^{+0.19}$ & $1.01_{-0.20}^{+0.20}$ & $1.02_{-0.22}^{+0.21}$ \\ 
$\psi$ & $0.93_{-0.13}^{+0.14}$ & $0.93_{-0.14}^{+0.15}$ & $0.94_{-0.14}^{+0.15}$ \\ 
$p_{\rm off}$ & $0.34_{-0.18}^{+0.18}$ & $0.37_{-0.21}^{+0.24}$ & $0.36_{-0.19}^{+0.19}$ \\ 
${\cal R}_{\rm off}$ & $2.2_{-1.3}^{+1.5}$ & $2.3_{-1.5}^{+1.5}$ & $2.4_{-1.4}^{+1.4}$ \\ 
$\alpha_{\rm inc}$ & $0.44_{-0.34}^{+0.35}$ & $0.53_{-0.24}^{+0.25}$ & $0.57_{-0.18}^{+0.20}$ \\ 
$\log M_{\rm inc}$ & $13.57_{-0.30}^{+0.32}$ & $13.88_{-0.30}^{+0.33}$ & $14.08_{-0.29}^{+0.32}$ \\ 
$\Omega_{\rm m}$ & $0.310_{-0.020}^{+0.019}$ & $0.306_{-0.020}^{+0.021}$ & $0.304_{-0.021}^{+0.022}$ \\ 
$\sigma_{8}$ & $0.785_{-0.044}^{+0.044}$ & $0.839_{-0.061}^{+0.061}$ & $0.813_{-0.059}^{+0.060}$ \\ 
$100\Omega_{\rm b}h^2$ & $2.228_{-0.040}^{+0.040}$ & $2.226_{-0.039}^{+0.040}$ & $2.222_{-0.041}^{+0.041}$ \\ 
$n_{\rm s}$ & $0.964_{-0.011}^{+0.011}$ & $0.963_{-0.010}^{+0.011}$ & $0.961_{-0.011}^{+0.011}$ \\ 
$h$ & $0.703_{-0.018}^{+0.018}$ & $0.700_{-0.016}^{+0.017}$ & $0.695_{-0.017}^{+0.017}$ \\ 

\hline
\hline
\end{tabular}
\end{center}
\medskip
\begin{minipage}{\hssize}
The three columns list the $68\%$ confidence intervals on the model parameters
for the three stellar mass subsamples we use in our analysis. The
parameter $M_{_*,11}$ denotes the stellar mass in units of
$10^{11}~h^{-2}M_\odot$ and the $68\%$ limit we quote is a one-sided
upper limit.
\end{minipage}
\end{table}

\renewcommand{\tabcolsep}{0.5cm}
\begin{table*}
\begin{center}
\caption{Population synthesis models average stellar masses}
\label{table:compare}
\begin{tabular}{cccc}
\hline\hline
\hline
& Subsample A & Subsample B & Subsample C \\
Model & $\avg{M_\ast}$ $(10^{11}\,h^{-2}M_\odot)$ & $\avg{M_\ast}$
$(10^{11}\,h^{-2}M_\odot)$ & $\avg{M_\ast}$ $(10^{11}\,h^{-2}M_\odot)$ \\
\hline
Portsmouth Passive Kroupa  & 1.14       & 1.51          & 1.79 \\
Portsmouth Star-forming Kroupa  & 0.97          & 1.28          & 1.53 \\
Portsmouth Passive Salpeter  & 1.95     & 2.55          & 2.96 \\
Portsmouth Star-forming Salpeter  & 1.51        & 1.99          & 2.38 \\
Granada Early-forming dust Kroupa  & 3.07       & 3.72          & 4.18 \\
Granada Late-forming dust Kroupa  & 2.59        & 3.11          & 3.46 \\
Granada Early-forming nodust Kroupa  & 2.58     & 3.13          & 3.53 \\
Granada Late-forming nodust Kroupa  & 2.19      & 2.62          & 2.95 \\
Granada Early-forming dust Salpeter  & 5.09     & 6.18          & 6.93 \\
Granada Late-forming dust Salpeter  & 4.33      & 5.19          & 5.79 \\
Granada Early-forming nodust Salpeter  & 4.33   & 5.25          & 5.93 \\
Granada Late-forming nodust Salpeter  & 3.67    & 4.41          & 4.96 \\
\hline
\hline
Our model & $<3.0$ (68\%) & $<7.4$ (68\%) & $<8.1$ (68\%) \\
\hline
\hline
\end{tabular}
\end{center}
\medskip
\begin{minipage}{2\hssize}
The average stellar mass estimate for the subsamples of galaxies used to measure
the clustering and lensing of galaxies in the present study from different SPS
models. The sample was defined using the first of these models, and the same
sample was used to estimate the average stellar mass in all cases. For
comparison, our $68\%$ model constraints are listed in the bottom two rows for
the fiducial and the off-centering model, respectively.
\end{minipage}
\end{table*}

The constraints on the halo occupation distribution for the three samples are
shown in Figure~\ref{fig:nofm}. As expected, the HOD shifts to higher halo
masses for subsamples B and C. 
The scatter in halo masses increases significantly for the
highest threshold sample. A constant scatter in stellar masses at fixed halo
mass translates into an increasing scatter in halo masses at fixed stellar mass
due to the shallow power law index of the stellar mass halo mass relation at
the massive end \citep[see e.g.,][]{More:2009b}. We compare our HOD constraints
to those obtained by \citet{White:2011} for the full sample of CMASS galaxies
in the left hand panel of Figure~\ref{fig:nofm}. Compared to their sample, our
fiducial subsample of galaxies resides in slightly larger halo masses owing to
our subsample selection which removes low stellar mass galaxies. Next we
compare the HOD of our subsample  C to that of luminous red
galaxies (LRGs) from SDSS-I/II at $z\sim0.3$. These two samples have comparable number
densities.  In the right hand panel, we compare the HOD constraints with those
obtained by \citet{Reid:2009} for luminous red galaxies shown with solid line.
The comparison demonstrates the similarity in the HOD of the two samples. For
reference, we also show the HOD of LRGs shifted to adjust for the difference in
the mass of the halos owing to the difference in the average redshift of the
LRG sample ($z\sim0.3$) and that of our subsample ($z\sim0.53$). This
particular HOD should hold if all LRGs have both maintained their identities
(central or satellite) inside their respective halos and if none of these halos
merged with each other in the redshift interval $z\in[0.35,0.57]$.

We also compare our HOD constraints with results obtained by
\citet{Leauthaud:2012} with a similar analysis of the abundance, clustering and
lensing signal but with data from the COSMOS survey. The gray shaded regions in
each of the panels of Figure~\ref{fig:nofm} show the 68 and 95 percent
confidence intervals obtained using the parameter constraints from
\citet{Leauthaud:2012}, for the same stellar mass cut as our selection in each
panel \footnote{We thank A. Leauthaud for providing us with
samples from the posterior distribution of their parameters. We have also
cross-compared the stellar masses of CMASS galaxies in our catalogs with those
from the COSMOS catalog and find no significant systematic biases between the
stellar mass determinations (apart from a scatter of $\sim$0.1 dex).}. The
COSMOS sample covers a much smaller area on the sky but is expected to be more
stellar mass complete compared to the CMASS sample, which has color cuts
designed to select the luminous red galaxy population.  The comparison shows
that the galaxies in the fiducial subsample on average reside in slightly larger
halo masses than the results from COSMOS. This result is consistent with the
expectation that massive redder galaxies reside in higher mass halos on average
\citep[see e.g.,][]{More:2011}. The results for subsamples B and C are, however,
consistent with the results from COSMOS, implying that the incompleteness in our
subsamples becomes smaller for the larger stellar mass thresholds. As the blue
fraction of galaxies decreases steeply as a function of stellar mass, the CMASS
colour selection no longer biases the sample. Tinker et al. (in preparation)
demonstrate this trend of decreasing incompleteness with increasing stellar mass
threshold in the CMASS sample based on galaxies that did not pass the CMASS
colour cuts but were observed as part of an SDSS-III ancillary program (J.
Tinker, priv. comm.).

The clustering and lensing signals of satellite galaxies are
different from that of central galaxies on small scales. Therefore it
is important to obtain a good estimate on the fraction of galaxies that are
satellites in our subsamples in order to correctly interpret the
measurements. The satellite fraction of our fiducial subsample is
$8.3\pm3.1$ percent. This value is consistent with the satellite fraction
quoted by \cite{White:2011}, $10\pm2$ percent, albeit on the lower
side. This is not entirely unexpected as our subsample excludes 
the low stellar mass galaxies in the entire CMASS sample that was
used by \cite{White:2011}. 
The satellite fraction is expected to decrease as a function of
stellar mass. The satellite fractions in our higher stellar mass threshold
subsamples are $6.8\pm2.6$ and $4.6\pm2.1$, respectively, consistent
with this expectation.

With the large flexibility in our modeling, we obtain very weak constraints on
the off-centering parameters, $p_{\rm off}$ and ${\cal R}_{\rm off}$. We also
do not detect any significant deviation of the concentration-mass relation from
the assumed relation calibrated from numerical simulations. We have observed
that when the off-centering parameters are not included, the posterior
distribution of ${\cal R}_{\rm c}$ shifts to lower values (by about 20
percent). A similar magnitude offset in the concentration-mass relation was
also observed by \cite{Mandelbaum:2008} by analyzing the weak lensing signal
(whilst ignoring the off-centering issue) on a wider range of mass scales, from
MaxBCG clusters \citep{Koester:2007}, LRGs, and $L_\ast$ lens
\citep{Mandelbaum:2006}.

The large-scale effective bias of galaxies (defined as in
Equation~\ref{eq:beff}) of our fiducial subsample of galaxies is constrained to
be $2.15\pm0.13$. This value is in agreement with the bias value of $2.05\pm0.3$
inferred by \citet{Comparat:2013} using the angular clustering and weak lensing
of all CMASS galaxies that were flagged as potential spectroscopic targets in
the CFHT Stripe 82 region\footnote{The small differences in the best fit bias
can be attributed to the different sample selection.}. The galaxy bias systematically increases to
$2.26\pm0.17$ and $2.54\pm0.19$, respectively, for subsamples with progressively
larger stellar mass thresholds. In the left hand panel of
Figure~\ref{fig:bias_crossr}, we show the scale dependence of the bias defined
as
\begin{equation}
b(r)=\left[\frac{\xi_{\rm gg}(r)}{\xi_{\rm mm}(r)}\right]^{1/2}\,,
\end{equation}
where $\xi_{\rm gg}$ and $\xi_{\rm mm}$ are the three dimensional galaxy and
matter correlation functions, respectively. The magenta, blue and yellow lines
enclose the 68 percent confidence intervals for our subsamples A, B and C after
marginalizing over all of our model parameters, respectively. The right hand
panel shows the cross-correlation coefficient defined as
\begin{equation}
r_{\rm ccc}(r)=\frac{\xi_{\rm gg}(r)}{\left[\xi_{\rm mm}(r)\xi_{\rm gm}(r)\right]^{1/2}}\,,
\end{equation}
where $\xi_{\rm gm}(r)$ is the three dimensional galaxy-matter
cross-correlation. This is a prediction based on the HOD models allowed by the
data. This prediction is consistent with the value of $r_{\rm
ccc}(r)=1.16\pm0.35$ measured by \citet{Comparat:2013} averaged on scales of
$0.1$--$2\mpch$. The cross-correlation coefficient is larger than one on small
scales \citep[see e.g.,][]{Seljak:2000}, but tends to unity on large scales.
This behaviour of the cross-correlation coefficient on large scales can be used
to obtain the matter correlation function directly from observations of the
galaxy clustering and galaxy-galaxy lensing on large scales \citep{Seljak:2000,
Guzik:2001, Baldauf:2010, Cacciato:2012, Mandelbaum:2013}.

In Figure~\ref{fig:Hinton_11_10}, we present the different degeneracies that are
inherent to our analysis. Large white (black) squares indicate positive
(negative) correlations in the inferred parameters. The scatter in halo masses,
$\sigma^2$, is tightly correlated with the mass scale, $M_{\rm min}$, above
which halos host one central galaxy. This degeneracy is expected due to the
dependence of each of the observables on these two parameters. Increasing
$M_{\rm min}$ results in increasing the mean halo mass of galaxies (and also the
galaxy bias relevant for the large scale clustering).  However, one can
compensate for this increase by increasing the scatter, and thereby including
more lower mass halos. The off-centering parameters are degenerate with each
other and show weak degeneracies with the satellite galaxy parameters and the
incompleteness parameters. In the case of a large off-centering fraction, we
expect the central signal to mimic the satellite signal which can explain the
degeneracies we observe. The cosmological parameters, $\Omega_\rmm$ and
$\sigma_8$ are degenerate with each other (see Section~\ref{sec:theory}), and
show degeneracies with the other cosmological parameters, which we will discuss
shortly.  However, more importantly, $\Omega_\rmm$ and $\sigma_8$ show only weak
degeneracies with other HOD parameters. This separation of degeneracies between
the cosmological parameters and the HOD parameters is a useful feature of the
joint analyses of the clustering and lensing of galaxies \citep{More:2013}.

We show the posterior distributions of our cosmological constraints as
histograms in the diagonal panels of Figure~\ref{fig:cosmo_degen} for
our fiducial subsample. The shaded regions highlight the degeneracies between
the cosmological parameters. The distributions shown by the blue solid lines
denote the priors that we have assumed on the auxiliary cosmological parameters
$\Omega_\rmb h^2, n_\rms$ and $h$ based on the analysis of WMAP9+SPT+ACT
\citep{Hinshaw:2013}. There is no significant improvement on these parameters
with the addition of our data, neither does our data shift the posteriors of
these parameters away from the priors \citep[cf.][]{Cacciato:2013}. We also
observe the degeneracy in the constraints on $\Omega_\rmm$ and $\sigma_8$ that
we anticipated from our theoretical considerations: a larger value of
$\Omega_\rmm$ prefers a model with smaller value of $\sigma_8$ as implied from
Figure~\ref{fig:toy}.

The cosmological constraints on $\Omega_\rmm$ and $\sigma_8$ obtained
from the three stellar mass subsamples analysed in this
paper are depicted in Figure~\ref{fig:Omm_s8_systematics} with
magenta, blue and yellow contours, respectively. The constraints from our
higher threshold subsamples, which are expected to be more complete, are
consistent with the fiducial subsample\footnote{A quantitative measure of the
consistency would require an analysis  of the clustering and lensing
measurements of all subsamples together including the correlation between the
subsamples. This investigation is beyond the scope of this work. The large overlap in the
68 and 95 percent confidence levels suggests there exists cosmological parameter
space which can simultaneously explain the measurements in each of the
subsample.}. Thus, although the fiducial subsample may have been incomplete,
this incompleteness does not cause significant biases in the cosmological
constraints, compared to the current statistical precision.  Given that the HOD
for each of the subsamples differs significantly from each other, this
agreement is non-trivial.

In our analysis, we have analysed the measurements of clustering to
$r_\rmp>0.85 \mpch$. We tested for additional systematics by extending our
measurements to even smaller scales $r_\rmp>0.5 \mpch$, but restrict
ourselves above the fiber collision scale at the highest redshift of our
subsamples. The small-scale clustering signal is sensitive to the satellite
fraction, and thus our constraints are expected to marginally improve. The 68
and 95 percent confidence levels on $\Omega_\rmm-\sigma_8$ for such an analysis
of our fiducial subsample are compared to those obtained when analyzing the
larger scale measurements in Figure~\ref{fig:rp_systematics}. This test also does not
reveal any significant biases.

The overlap between the CFHTLS region and the BOSS region consists of
only about 105 deg$^2$. Therefore, sample variance for the lensing signal is
a legitimate concern for our lensing analysis. The CFHTLS
region consists of four different fields which are well separated from one
another. The effect of super-survey modes that can affect the measurement is
smaller in this case than when the survey area is contiguous
\citep{Takada_2013}. Regardless, if the CFHTLS region happens to represent a
relatively under or over-dense part of the Universe, then there are a number of
ways in which our analysis could be affected. 

First, the number density of galaxies in the particular patch could be affected.
This change has a relatively minor effect on the lensing signal, since the
signal is normalized by the total number of galaxies (see the factors ${\cal
H}_{\rm x}$ in Equations \ref{P1h}-\ref{P2h}). Second, the concentration of
halos can depend upon the environment \citep[see e.g.,][]{Maccio:2007}, so
the halos of the lensing galaxies could have density profiles which deviate from
their expected median. In this case, the nuisance parameter, $\cal R_\rmc$, that
we adopt should be able to marginalize over this uncertainty. The third
possibility is that the two-halo term of the lensing contribution is affected in
these regions \citep[see e.g.,][]{Gao:2007}. To explore the impact of this
possibility, we removed all lensing information from scales above $2.6 \mpch$
for our fiducial subsample where the two-halo term is expected to be dominant.
The resulting cosmological constraints and their comparison with our fiducial
analysis is shown in Figure~\ref{fig:esdmax}. As expected, the errorbars are
larger when we restrict our analysis to small scales.  Although there is a
slight tendency toward larger $\Omega_\rmm$ and $\sigma_8$ values, both the 68
and the 95 percent confidence levels overlap to a large extent. Finally for
completeness, we mention that it could be possible that galaxy formation is
heavily dependent on the local cosmological parameters, and this could affect
our analysis. However, a proper study of this issue is beyond the scope of this
paper. All of these effects can be remedied by surveying larger portions of the
sky.

In our analysis, we have ignored the cross-covariance between the clustering and
the weak lensing signal. Our weak lensing analysis is restricted to the small
overlap area between the CFHTLS and BOSS survey, which justifies our assumption
of setting the cross-covariance of the signals to zero. In addition, the
presence of shape noise tends to reduce the cross-covariance between the
clustering and the lensing signal. Nevertheless, we have
repeated our entire analysis with the clustering signal obtained by excluding
galaxies in the CFHTLS regions. We confirm that none of our results are
affected by the ignorance of the cross-covariance between the clustering and
lensing signals.

In Figure~\ref{fig:Omm_s8}, we compare the cosmological constraints on
$\Omega_\rmm$ and $\sigma_8$ from our fiducial subsample with the results
obtained by a variety of other complementary methods. The 68 and 95 percent
confidence intervals obtained by the CMB temperature fluctuation power spectrum
measurements of WMAP9 \citep{Hinshaw:2013} in combination with the high-multipole measurements of the
same from SPT \citep{Keisler:2011} and ACT \citep{Das:2011} are shown as green shaded regions and denoted as WMAP9\_E.
The chrome yellow shaded regions show the confidence intervals obtained by the
Planck collaboration using the temperature power spectrum measurements \citep{Planck:2013xvi}. 
The gray bands correspond to the 68 and 95 percent confidence
constraints obtained by Planck \citep{Planck:2013xxi} but using the thermal Sunyaev-Zel'dovich
(SZ) power
spectrum measurements \citep[][for the SZ effect]{Sunyaev:1972},
 while the brown shaded regions denote the constraints
obtained from Planck SZ cluster abundances \citep{Planck:2013xx}. The confidence contours obtained by
performing a joint analysis of the abundance, clustering and lensing 
signals of the SDSS main sample of galaxies carried out by
\citet{Cacciato:2013} are shown using dark blue contours, while the
constraints obtained by \citet{Mandelbaum:2013}, using a joint analysis
of clustering and lensing but focusing on large scales, are shown as
yellow contours. The dark purple shaded regions correspond to the
analysis of the tomographic weak lensing signal from CFHTLenS by
\cite{Heymans:2013}. The red shaded confidence regions are the
results of \cite{Beutler:2013}, obtained by combining the baryon
acoustic oscillation measurements (assuming the Planck value for the
sound horizon)\footnote{The contours shift to the right by  $\sim
2-\sigma$ if the value of the sound horizon from WMAP9 is assumed.},
the redshift space distortions and the Alcock-Paczynski test 
\citep{Alcock:1979} from the
CMASS galaxy sample. 

The two results from the cosmic microwave background (WMAP9\_E and Planck) are
in agreement, although there is a noticeable difference in the
central values obtained from the analysis by the two teams. The WMAP9\_E
analysis prefers a lower value for both $\Omega_\rmm$ and $\sigma_8$ compared to
the Planck analysis. Our confidence regions overlap with both WMAP9\_E and
Planck, and roughly lie in an orthogonal direction. This results demonstrates
that our constraints are complementary to those obtained from the CMB analyses. 

The measurements from the thermal SZ power spectrum and the SZ cluster
abundances from Planck are also consistent with each other and with the WMAP9\_E
and Planck CMB analysis. However, at the central value of $\Omega_\rmm$
preferred by Planck, both these analyses prefer a much lower value of
$\sigma_8$. The same tendency for the preference of a lower value of $\sigma_8$
at the central value of $\Omega_\rmm$ preferred by Planck is also seen in our
results as well as those from the SDSS conditional luminosity function (CLF)
analysis of \citet{Cacciato:2013}, LRG clustering and weak lensing analysis of
\citet{Mandelbaum:2013}, the CFHTLenS tomographic weak lensing analysis of
\citet{Heymans:2013} and the redshift space distortion (RSD)
measurements of \citet{Beutler:2013}. However, there is a common region of
overlap between the different analyses: this common region suggests slightly
smaller (larger) values of both $\Omega_\rmm$ and $\sigma_8$ compared to Planck
(WMAP9\_E) CMB constraints. Interestingly, the results from a reanalysis of the
Planck data using a different foreground cleaning procedure performed by
\citet[][shown using light blue shaded regions]{Spergel:2013} also results in constraints
in the same overlapping region. The more recent BAO analyses also hint towards
an intermediate value for $\Omega_\rmm$ in between WMAP9 and Planck
\citep{Anderson:2014}. Our results are also consistent with constraints
from the cross-correlation of the thermal Sunyaev Z\'eldovich
signal from Planck with the gravitational lensing potential \citep{Hill:2014}
and with the X-ray cluster map from ROSAT \citep{Hajian:2013}, both of these
analyses also prefer intermediate values of the parameters $\Omega_\rmm$ and
$\sigma_8$.

In Figure \ref{fig:fsigma8}, we present constraints on $f\sigma_8(z)$, where
$f=-{\rm d}\ln{D(z)}/{\rm d}\ln(1+z) $ is the logarithmic growth rate and
$\sigma_8(z)=\sigma_8\,D(z)/D(0)$ is the linear matter fluctuation at redshift
$z$, that are calculated from our measurement and other joint analyses of
clustering and lensing measurements \citep{Cacciato:2013, Mandelbaum:2013}. We
also show the confidence regions for a $\Lambda$CDM model assuming cosmological
parameters from WMAP9\_E and Planck using green and chrome yellow shaded bands.
Our measurement is at the highest redshift among the clustering and lensing
joint analyses and is consistent with these measurements and the WMAP9\_E and
Planck predictions. For comparison, we have also compiled various RSD 
measurements \citep{Percival:2004, Blake:2011b, Samushia:2012, Beutler:2012,
Chuang:2013a, de_la_Torre:2013, Chuang:2013b, Beutler:2013, Samushia:2014,
Reid:2014}.  Our measurement is also largely consistent with the RSD
measurements.

In our modeling exercise we have assumed that the halo
mass of a galaxy has the dominant effect in determining its properties. Our halo
occupation distribution formalism assumes that the halos of a given
mass which host the galaxies from our subsamples are a random subsample of halos
of that particular mass. It assumes that the presence or absence of a galaxy is
not determined by the assembly history of the halo. The extent to which this
assumption holds is, however, unclear. It is important to note that our parent
galaxy sample is designed to select galaxies based on colour. Recently, subhalo
abundance matching methods have been extended to assign both stellar mass and
colours to galaxies in mock catalogs \citep{Hearin:2013a}. Their methods are based on the
simple idea that properties of galaxies such as their colour or star formation
rate may depend on the formation age of the halo defined in a suitable manner.
These models have been successfully employed to qualitatively match the
colour-dependent clustering of galaxies and the galaxy-galaxy lensing signal
\citep{Hearin:2013b}. Such mock galaxy catalogs by construction have assembly
bias. In such cases, halo age in addition to the mass decides the colour of
galaxies. 

If such models reflect the true nature of galaxy formation in the
Universe, then the selection applied to determine our parent galaxy sample
(CMASS selection) may identify galaxies at a given fixed halo mass that
preferentially live in halos which formed earlier \citep[see
e.g.,][for a discussion of the effect
in the SDSS main galaxy sample]{Zentner:2013}. It is well known that the clustering
of halos at fixed halo mass depends upon the formation age of the halo: halos
that form earlier cluster more strongly than average \citep{Gao:2007}, which
can be problematic for our inference of cosmological parameters from the halo
mass-bias relation. However, it is also known that the difference in the
formation time-dependent clustering of halos is less pronounced at the high
mass end. For halos that form from the initial peaks with height
$\nu>1.8$, the formation age dependence is negligible. The
fiducial subsample used in our analysis has a bias $b=2.15\pm0.13$, which
corresponds to peak heights $\nu\sim2.0$. The high value of peak height limits
the impact that assembly bias can have on our results. Nevertheless, a more
thorough study of how assembly bias properties other than the formation
history, e.g. the spin of halos, their concentrations, etc. can affect our
results is warranted.

In addition, the theoretical foundation of our results are the calibrations of
the matter density distributions from collisionless numerical simulations.
Baryonic processes such as radiation pressure feedback, feedback from
supernovae and active galactic nuclei, which are implemented in large volume
hydrodynamical simulations, can have a significant effect on the matter power
spectrum, the halo mass function and the halo bias functions
\citep{Gnedinetal:04,Rudd:2008, Cui:2012,Velliscig:2014,vanDaalen:2014,Vogelsberger:2014}.
These consequences are a result of the redistribution of matter in and around
halos due to the baryonic feedback effects. The magnitude of the difference is
however very dependent on the details of the feedback mechanisms implemented.
Our inclusion of the weak prior on the amplitude of the concentration-mass
relation (instead of complete reliance on the simulation calibrated
normalization) can partially account for a variation in the matter density
within halos that could be a result of baryonic effects \citep[see
e.g.,][]{Zentner:2013b}. Nevertheless, the
impact of baryonic effects is certain to remain a subject of active research in
the near future. 

\section{Summary and future outlook}
\label{sec:conc}

Galaxies are biased tracers of the matter density distribution. Galaxy bias is
often treated as a nuisance parameter to be marginalized over in order to use
the two-point functions of galaxies to derive cosmological parameters. The
origin of galaxy bias is, however, in the special positions that galaxies occupy
in the matter density field. Galaxies form within halos which are located at the
peaks of the density distribution. They share the bias of the halos in which
they reside. The dependence of halo bias on mass is governed by cosmological
parameters. Therefore, measurements of the clustering amplitude of
galaxies,
 in
combination with the halo masses of galaxies, can turn the nuisance of galaxy
bias into a powerful probe of cosmological parameters. In this paper,
we utilized such measurements in order to constrain the halo occupation distribution
of galaxies as well as the cosmological parameters $\Omega_\rmm$ and
$\sigma_8$.

For this purpose, we used spectroscopic galaxies from the SDSS-III BOSS project
which span an area of about 8500 deg$^2$ in the sky. From this parent sample,
we constructed a subsample of galaxies so that it obeyed stellar mass limits
($\log M_*/h_{70}^{-2} M_\odot \in [11.10,12.0]$) and was approximately complete
within the redshift range $z\in[0.47,0.59]$, with an approximately constant
abundance. This subsample of galaxies was used to measure the projected galaxy
clustering signal with a signal-to-noise ratio of $56$ for scales
$0.85\mpch<r_\rmp<80.0\mpch$.  We made use of the
publicly available galaxy shape and photometric redshift catalogs compiled by the CFHTLenS collaboration
based on deeper, higher quality imaging data from the CFHTLS. This imaging
catalog had an overlap of a mere 105 deg$^2$  with the BOSS footprint, but it
allowed measurement of the weak gravitational lensing signal of BOSS galaxies
with a signal-to-noise ratio of $26$ for scales
$0.1\mpch<r_\rmp<20.0\mpch$.  To test for systematics arising from our sample
selection we also measured the clustering and lensing signals for two other
subsamples within the same redshift range, but with larger thresholds in
stellar mass $\log M_*/h_{70}^{-2} M_\odot \in [11.30,12.0]$ and $\log
M_*/h_{70}^{-2} M_\odot \in [11.40,12.0]$, respectively.

We analyzed these measurements in the framework of the halo model. Our halo
model uses a number of ingredients for which the cosmological dependence has
been calibrated using numerical simulations, e.g., the halo mass function, the
halo bias function, the density profile of halos, the radial dependence of the
halo bias. We also utilize well-tested prescriptions to implement halo exclusion
and correct for residual redshift space distortion effects in the projected
clustering signal due to the use of finite line-of-sight integration limit while
projecting the clustering signal. In addition, we also allow for a baryonic
component at the center of halos, use parameters to describe the potential
incompleteness in the sample, as well as parameters which allow a fraction of
central galaxies to be offset from the true center of the halo. We performed an
MCMC analysis to obtain the posterior distributions of all our model parameters
of interest given our measurements and after marginalizing over all of the
nuisance parameters. Our analytical model consists of 17 parameters
in total, $5$ describing the halo occupation distribution of galaxies, $1$ for
the stellar mass contribution, $1$ for the concentration-mass relation
normalization, $5$ nuisance parameters, the cosmological parameters,
$\Omega_bh^2,n_\rms,h$ with WMAP9+SPT+ACT priors, and $\Omega_\rmm$ and
$\sigma_8$ were left completely free.

Our model is successful in reproducing the abundance, the
projected clustering and the galaxy-galaxy lensing signal in our fiducial
subsample as well as the larger threshold stellar mass subsamples. We obtained
constraints on the halo occupation distribution parameters of galaxies in each
of our subsamples and the cosmological parameters $\Omega_\rmm$ and
$\sigma_8$.  We compared the HOD constraints from our
analysis with those obtained by \citet{Leauthaud:2012} and found that our
fiducial subsample may be biased towards high mass halos at the
stellar mass threshold of our subsample. However, this bias declines
substantially for the larger threshold subsamples. Nevertheless, the
cosmological constraints from the analysis of each of the subsamples are in
agreement with each other, and reveal no significant biases in the
cosmological parameter estimates given the current errors. 

The cosmological constraints from the analysis of our fiducial subsample yield
$\Omega_\rmm=0.310^{+0.019}_{-0.020}$ and $\sigma_8=0.785\pm0.044$. This is in
excellent agreement with constraints obtained by a number of different studies,
including CMB temperature fluctuation power spectrum measurements from
WMAP9+SPT+ACT, Planck, and other independent constraints from SZ cluster
abundances, SZ thermal power spectrum measurements, cosmic shear measurements
and baryon acoustic oscillation measurements combined with redshift space
distortions and the Alcock Paczynski test. Furthermore, our results are also
consistent with those obtained by a joint analysis of the clustering and
lensing of galaxies of the SDSS main sample of galaxies and those of LRGs.

Our analysis extends the redshift at which cosmological constraints
have been obtained using a joint clustering and lensing analysis to
$z=0.53$. In the near future, the Subaru Hyper Suprime-Cam survey,
which began in the spring of 2014, is expected to provide deeper and
better quality imaging in the SDSS-III BOSS footprint, extending the
overlap region by more than an order of magnitude to an unprecedented
1400 deg$^2$. This will allow a more detailed study of the galaxy-dark
matter connection of the BOSS galaxies, and allow division of the
sample into finer redshift and stellar mass bins. Analyses such as these
with a larger redshift lever arm have the potential to provide
complementary and competitive constraints on cosmological models with
an extended parameter set such as those with an evolving dark energy
equation of state \citep[e.g.,][]{OguriTakada:11}.

\section*{Acknowledgments}

We thank the referee for detailed and useful comments which greatly improved the
readability of both Paper I and II.
SM and MT were supported by World Premier International Research Center
Initiative (WPI Initiative), MEXT, Japan, by the FIRST program ``Subaru
Measurements of Images and Redshifts (SuMIRe)'', CSTP, Japan. HM
was supported by Japan Society for the Promotion of Science
(JSPS) Postdoctoral Fellowships for Research Abroad and JSPS Research
Fellowships for Young Scientists. RM was supported by the Department of
Energy Early Career Award program.  MT was supported by Grant-in-Aid for
Scientific Research from the JSPS Promotion of Science (No.~23340061 and
26610058).  DNS acknowledges support from NSF grant AST1311756 and NASA
ATP grant NNX12AG72G.  SM would like to thank Naoshi Sugiyama for kindly
allowing the use of the computing cluster COSMOS at the Physics
Department of the University of Nagoya for the analysis presented in
this paper. SM is also grateful to Saga Shohei for the administrative
support in this regard. The authors would like to thank Ramin Skibba, David
Hogg, Marcello Cacciato, Johan Comparat, Sarah Bridle and Catherine Heymans for
useful comments on an earlier version of the paper.

Funding for SDSS-III has been provided by the Alfred P. Sloan Foundation, the Participating Institutions, the National Science Foundation, and the U.S. Department of Energy Office of Science. The SDSS-III web site is http://www.sdss3.org/.

SDSS-III is managed by the Astrophysical Research Consortium for the
Participating Institutions of the SDSS-III Collaboration including the
University of Arizona, the Brazilian Participation Group, Brookhaven
National Laboratory, University of Cambridge, Carnegie Mellon
University, University of Florida, the French Participation Group, the
German Participation Group, Harvard University, the Instituto de
Astrofisica de Canarias, the Michigan State/Notre Dame/JINA
Participation Group, Johns Hopkins University, Lawrence Berkeley
National Laboratory, Max Planck Institute for Astrophysics, Max Planck
Institute for Extraterrestrial Physics, New Mexico State University, New
York University, Ohio State University, Pennsylvania State University,
University of Portsmouth, Princeton University, the Spanish
Participation Group, University of Tokyo, University of Utah, Vanderbilt
University, University of Virginia, University of Washington, and Yale
University.

This work is based on observations obtained with MegaPrime/MegaCam, a joint
project of CFHT and CEA/IRFU, at the Canada-France-Hawaii Telescope (CFHT)
which is operated by the National Research Council (NRC) of Canada, the
Institut National des Sciences de l'Univers of the Centre National de la
Recherche Scientifique (CNRS) of France, and the University of Hawaii. This
research used the facilities of the Canadian Astronomy Data Centre operated by
the National Research Council of Canada with the support of the Canadian Space
Agency. CFHTLenS data processing was made possible thanks to significant
computing support from the NSERC Research Tools and Instruments grant program.


\bibliographystyle{apj}
\bibliography{paper}

\end{document}